\documentclass[twocolumn]{aastex63}

\usepackage{multirow}
\usepackage{amsmath}
\usepackage{graphicx}
\usepackage{subfigure}
\usepackage{savesym}
\savesymbol{tablenum}
\usepackage{siunitx}
\restoresymbol{SIX}{tablenum}

\usepackage{svg}
\usepackage{xspace}
\usepackage{stackengine}
\usepackage{soul}
\usepackage{txfonts}
\usepackage[switch]{lineno}
\def\newacronym#1#2#3{\gdef#1{#3 (#2)\gdef#1{#2}}}
\usepackage{subfiles} 

\begin{document}

\newacronym{\dr}{DR}{dilution refrigerator}
\newacronym{\mc}{MC}{mixing chamber}


   \title{The Simons Observatory: Design, integration, and testing of the small aperture telescopes}
   \author[0000-0001-7225-6679]{Nicholas Galitzki}
    \affiliation{Department of Physics, University of Texas at Austin, Austin, TX, 78712, USA}
    \affiliation{Weinberg Institute for Theoretical Physics, Texas Center for Cosmology and Astroparticle Physics, Austin, TX 78712, USA}
    \author[0000-0002-1667-2544]{Tran Tsan}
    \affiliation{Department of Physics, University of California San Diego, San Diego, CA 92093, USA}
    \author{Jake Spisak}
    \affiliation{Department of Physics, University of California San Diego, San Diego, CA 92093, USA}
    \author{Michael Randall}
    \affiliation{Department of Physics, University of California San Diego, San Diego, CA 92093, USA}
    \author[0000-0001-7480-4341]{Max Silva-Feaver}
    \affiliation{Department of Physics, University of California San Diego, San Diego, CA 92093, USA}
    \author{Joseph Seibert}
    \affiliation{Department of Physics, University of California San Diego, San Diego, CA 92093, USA}
    \author[0000-0002-6522-6284]{Jacob Lashner}
    \affiliation{Wright Laboratory, Department of Physics, Yale University, New Haven, CT 06520, USA}

    \author[0000-0002-0400-7555]{Shunsuke Adachi}
        \affiliation{Hakubi Center for Advanced Research, Kyoto University, Kyoto 606-8501, Japan}
    \author[0009-0004-3835-600X]{Sean M. Adkins}
        \affiliation{Wave2 Enterprises, Kailua Kona, HI 96740, USA}
    \author[0000-0003-1942-1334]{Thomas Alford}
        \affiliation{Department of Physics, University of Chicago, Chicago, IL  60637, USA}
    \author[0000-0002-3407-5305]{Kam Arnold}
        \affiliation{Department of Physics, University of California San Diego, San Diego, CA 92093, USA}
        \affiliation{Department of Astronomy \& Astrophysics, University of California San Diego, San Diego, CA 92093, USA}
    \author[0009-0005-1673-0504]{Peter C. Ashton}
        \affiliation{SRI International, Menlo Park, CA 94025, USA}
    \author[0000-0002-6338-0069]{Jason E. Austermann}
        \affiliation{Quantum Sensors Division, NIST, Boulder, CO 80305, USA}
    \author[0000-0002-8211-1630]{Carlo Baccigalupi}
        \affiliation{The International School for Advanced Studies (SISSA), via Bonomea 265, I-34136 Trieste, Italy}
        \affiliation{The National Institute for Nuclear Physics (INFN), via Valerio 2, I-34127, Trieste, Italy}
        \affiliation{The Institute for Fundamental Physics of the Universe (IFPU), Via Beirut 2, I-34151, Trieste, Italy }
    \author[0000-0002-7888-6222]{Andrew Bazarko}
        \affiliation{Joseph Henry Laboratories of Physics, Jadwin Hall, Princeton University, Princeton, NJ 08544, USA}
    \author[0000-0003-1263-6738]{James A. Beall}
        \affiliation{Quantum Sensors Division, NIST, Boulder, CO 80305, USA}
    \author{Sanah Bhimani}
        \affiliation{Wright Laboratory, Department of Physics, Yale University, New Haven, CT 06520, USA}
    \author[0009-0008-4312-6814]{Bryce Bixler}
        \affiliation{Department of Physics, University of California San Diego, San Diego, CA 92093, USA}
    \author{Gabriele Coppi}
        \affiliation{Department of Physics, University of Milano - Bicocca, Piazza della Scienza, 3 - 20126 Milano, Italy}
    \author{Lance Corbett}
        \affiliation{Department of Physics, University of California - Berkeley, Berkeley, CA 94720, USA}
        \affiliation{Physics Division, Lawrence Berkeley National Laboratory, Berkeley, CA 94720, USA}
    \author[0000-0002-8351-3711]{Kevin D. Crowley}
        \affiliation{Joseph Henry Laboratories of Physics, Jadwin Hall, Princeton University, Princeton, NJ 08544, USA}
    \author[0000-0001-5068-1295]{Kevin T. Crowley}
        \affiliation{Department of Astronomy \& Astrophysics, University of California San Diego, San Diego, CA 92093}    
    \author[0009-0003-5814-2087]{Samuel Day-Weiss}
        \affiliation{Joseph Henry Laboratories of Physics, Jadwin Hall, Princeton University, Princeton, NJ 08544, USA}
    \author{Simon Dicker}
        \affiliation{Department of Physics and Astronomy, University of Pennsylvania, Philadelphia PA, 19104, USA}
    \author{Peter N. Dow}
        \affiliation{Department of Astronomy, University of Virginia, Charlottesville, VA 22904, USA}
    \author[0000-0002-6318-1924]{Cody J. Duell}
        \affiliation{Department of Physics, Cornell University, Ithaca, NY 14853, USA}
    \author[0000-0002-9693-4478]{Shannon M. Duff}
        \affiliation{Quantum Sensors Division, NIST, Boulder, CO 80305, USA}
    \author{Remington G. Gerras}
        \affiliation{Department of Physics and Astronomy, University of Southern California, Los Angeles, CA 90089, USA}
    \author[0000-0001-9880-3634]{John C. Groh}
        \affiliation{Physics Division, Lawrence Berkeley National Laboratory, Berkeley, CA, 94702, USA}
    \author[0000-0003-1760-0355]{Jon E. Gudmundsson}
        \affiliation{The Oskar Klein Centre, Department of Physics, Stockholm University, Stockholm, Sweden}
        \affiliation{Science Institute, University of Iceland, 107 Reykjavik, Iceland}
    \author[0000-0003-1248-9563]{Kathleen Harrington}
        \affiliation{High Energy Physics Division, Argonne National Laboratory, Lemont, IL 60439, USA}
        \affiliation{Department of Astronomy and Astrophysics, University of Chicago, Chicago, IL 60637, USA}
    \author[0000-0003-1443-1082]{Masaya Hasegawa}
        \affiliation{High Energy Accelerator Research Organization (KEK), Tsukuba, Ibaraki 305-0801, Japan}
    \author[0000-0002-3757-4898]{Erin Healy}
        \affiliation{Kavli Institute for Cosmological Physics, University of Chicago, Chicago, IL 60637, USA}
    \author[0000-0001-7878-4229]{Shawn W. Henderson}
        \affiliation{Kavli Institute for Particle Astrophysics and Cosmology, Stanford, CA 94305, USA}
        \affiliation{SLAC National Accelerator Laboratory, Menlo Park, CA 94025, USA}
    \author{Johannes Hubmayr}
        \affiliation{Quantum Sensors Division, NIST, Boulder, CO 80305, USA}
    \author{Jeffrey Iuliano}
        \affiliation{Department of Physics, Villanova University, 800 E. Lancaster Ave., Villanova, PA 19085, US}
    \author[0000-0002-6898-8938]{Bradley R. Johnson}
        \affiliation{Department of Astronomy, University of Virginia, Charlottesville, VA 22904, USA}
    \author{Brian Keating }
        \affiliation{Department of Physics, University of California San Diego, San Diego, CA 92093, USA}
    \author[0000-0002-2978-7957]{Ben Keller}
        \affiliation{Department of Physics, Cornell University, Ithaca, NY 14853, USA}
    \author{Kenji Kiuchi}
         \affiliation{Department of Physics, The University of Tokyo, Tokyo 113-0033, Japan}
    \author[0000-0001-5374-1767]{Anna M. Kofman}
        \affiliation{Department of Physics and Astronomy, University of Pennsylvania, Philadelphia PA, 19104, USA}
    \author[0000-0003-0744-2808]{Brian J. Koopman}
        \affiliation{Wright Laboratory, Department of Physics, Yale University, New Haven, CT 06520, USA}
    \author[0000-0003-3106-3218]{Akito Kusaka}
        \affiliation{Department of Physics, The University of Tokyo, Tokyo 113-0033, Japan}
        \affiliation{Physics Division, Lawrence Berkeley National Laboratory, Berkeley, CA 94720, USA}
        \affiliation{Research Center for the Early Universe, School of Science, The University of Tokyo, Tokyo 113-0033, Japan}
        \affiliation{Kavli Institute for the Physics and Mathematics of the Universe (WPI), UTIAS, The University of Tokyo, Chiba 277-8583, Japan}
    \author[0000-0003-3106-3218]{Adrian T. Lee}
        \affiliation{Department of Physics, University of California - Berkeley, Berkeley, CA 94720, USA}
        \affiliation{Physics Division, Lawrence Berkeley National Laboratory, Berkeley, CA 94720, USA}
    \author[0009-0009-0489-8720]{Richard A. Lew}
        \affiliation{Theiss Research, La Jolla, CA 92037, USA}
        \affiliation{Quantum Sensors Division, NIST, Boulder, CO 80305, USA}
    \author[0000-0001-5465-8973]{Lawrence T. Lin}
        \affiliation{Department of Astronomy, Cornell University, Ithaca, NY 14853, USA}
    \author{Michael J Link}
        \affiliation{Quantum Sensors Division, NIST, Boulder, CO 80305, USA}
    \author[0000-0001-7694-1999]{Tammy J. Lucas}
        \affiliation{Quantum Sensors Division, NIST, Boulder, CO 80305, USA}
    \author{Marius Lungu}
        \affiliation{Department of Physics and Astronomy, University of Pennsylvania, 209 S. 33rd St, Philadelphia PA, 19104, USA}
    \author[0009-0000-1028-3524]{Aashrita Mangu}
        \affiliation{Department of Physics, University of California - Berkeley, Berkeley, CA 94720, USA}
    \author{Jeffrey J McMahon}
        \affiliation{Department of Astronomy and Astrophysics, University of Chicago, Chicago, IL 60637, USA}
        \affiliation{Kavli Institute for Cosmological Physics, University of Chicago, Chicago, IL 60637, USA}
        \affiliation{Enrico Fermi Institute, University of Chicago, Chicago, IL  60637, USA}
        \affiliation{Department of Physics, University of Chicago, Chicago, IL  60637, USA}
        \affiliation{Fermi National Accelerator Laboratory, Batavia, IL 60510, USA}
    \author{Amber D. Miller}
        \affiliation{Department of Physics and Astronomy, University of Southern California, Los Angeles, CA 90089, USA}
    \author[0000-0002-7340-9291]{Jenna E. Moore}
        \affiliation{School of Earth and Space Exploration, Arizona State University, Tempe, AZ, 85287, USA}
    \author[0000-0002-3214-8881]{Magdy Morshed}
        \affiliation{Universit\'e Paris Cit\'e, CNRS, Astroparticule et Cosmologie, F-75013 Paris, France}
        \affiliation{CNRS-UCB International Research Laboratory, Centre Pierre Binétruy, IRL2007, CPB-IN2P3, Berkeley, US}
    \author[0000-0002-6300-1495]{Hironobu Nakata}
        \affiliation{Department of Physics, Faculty of Science, Kyoto University, Kyoto 606-8502, Japan}
    \author[0000-0002-8307-5088]{Federico Nati}
        \affiliation{Department of Physics, University of Milano - Bicocca, Piazza della Scienza, 3 - 20126 Milano, Italy}
    \author[0000-0002-7333-5552]{Laura B. Newburgh}
        \affiliation{Wright Laboratory, Department of Physics, Yale University, New Haven, CT 06520, USA}
    \author[0000-0002-7575-8145]{David V. Nguyen}
        \affiliation{Wright Laboratory, Department of Physics, Yale University, New Haven, CT 06520, USA}
    \author{Michael D. Niemack}
        \affiliation{Department of Physics, Cornell University, Ithaca, NY 14853, USA}
        \affiliation{Department of Astronomy, Cornell University, Ithaca, NY 14853, USA}
    \author[0000-0002-9828-3525]{Lyman A. Page}
        \affiliation{Joseph Henry Laboratories of Physics, Jadwin Hall, Princeton University, Princeton, NJ 08544, USA}
    \author[0000-0001-5667-8118]{Kana Sakaguri}
        \affiliation{Department of Physics, The University of Tokyo, Tokyo 113-0033, Japan}
    \author[0000-0001-6389-0117]{Yuki Sakurai}
        \affiliation{Faculty of Engineering, Suwa University of Science, Nagano, 391-0292, Japan}
        \affiliation{Kavli Institute for the Physics and Mathematics of the Universe (WPI), UTIAS, The University of Tokyo, Chiba 277-8583, Japan}
    \author[0000-0002-9761-3676]{Mayuri Sathyanarayana Rao}
        \affiliation{Raman Research Institute, Bengaluru, Karnataka 560080, India}
    \author[0000-0001-6367-6380]{Lauren J. Saunders}
        \affiliation{Fermi National Accelerator Laboratory, Batavia, IL 60510, USA}
    \author[0000-0003-0514-9034]{Jordan E. Shroyer}
        \affiliation{Department of Astronomy, University of Virginia, Charlottesville, VA 22904, USA}
    \author[0009-0007-7435-9082]{Junna Sugiyama}
        \affiliation{Department of Physics, The University of Tokyo, Tokyo 113-0033, Japan}
    \author[0000-0003-2439-2611]{Osamu Tajima}
        \affiliation{Department of Physics, Faculty of Science, Kyoto University, Kyoto 606-8502, Japan}
    \author{Atsuto Takeuchi}
        \affiliation{Department of Physics, The University of Tokyo, Tokyo 113-0033, Japan}
    \author[0000-0003-3264-5228]{Refilwe Tanah Bua}
        \affiliation{Department of Physics, Brown University, Providence, RI, 02912, USA}
    \author{Grant Teply}
        \affiliation{Department of Physics, University of California San Diego, San Diego, CA 92093, USA}
    \author[0000-0002-2380-0436]{Tomoki Terasaki}
        \affiliation{Department of Physics, The University of Tokyo, Tokyo 113-0033, Japan}
    \author[0000-0003-2486-4025]{Joel N. Ullom}
        \affiliation{Quantum Sensors Division, NIST, Boulder, CO 80305, USA}
    \author{Jeffrey L. Van Lanen}
        \affiliation{Quantum Sensors Division, NIST, Boulder, CO 80305, USA}
    \author[0000-0002-2105-7589]{Eve M. Vavagiakis}
        \affiliation{Department of Physics, Cornell University, Ithaca, NY 14853, USA}
    \author[0000-0003-2467-7801]{Michael R Vissers}
        \affiliation{Quantum Sensors Division, NIST, Boulder, CO 80305, USA}
    \author{Liam Walters}    
        \affiliation{Department of Astronomy, University of Virginia, Charlottesville, VA 22904, USA}
    \author[0000-0002-8710-0914]{Yuhan Wang}
        \affiliation{Joseph Henry Laboratories of Physics, Jadwin Hall, Princeton University, Princeton, NJ 08544, USA}
    \author[0000-0001-5112-2567]{Zhilei Xu}
        \affiliation{MIT Kavli Institute, Massachusetts Institute of Technology, Cambridge, MA 02139, USA}
    \author[0000-0003-0221-2130]{Kyohei Yamada}
        \affiliation{Joseph Henry Laboratories of Physics, Jadwin Hall, Princeton University, Princeton, NJ 08544, USA}
        \affiliation{Department of Physics, The University of Tokyo, Tokyo 113-0033, Japan}
    \author{Kaiwen Zheng}
        \affiliation{Joseph Henry Laboratories of Physics, Jadwin Hall, Princeton University, Princeton, NJ 08544, USA}

  \begin{abstract}
   { The Simons Observatory (SO) is a cosmic microwave background (CMB) survey experiment that includes small-aperture telescopes (SATs) observing from an altitude of 5,200\;m in the Atacama Desert in Chile. The SO SATs will cover six spectral bands between 27 and 280\;GHz to search for primordial B-modes to a sensitivity of $\sigma(r)=0.002$, with quantified systematic errors well below this value.  Each SAT is a self-contained cryogenic telescope with a 35$^\circ$ field of view, 42\,cm diameter optical aperture,  40\,K half-wave plate, 1\,K refractive optics, and $<0.1$\,K focal plane that holds $>12,000$ TES detectors. We describe the nominal design of the SATs and present details about the integration and testing for one operating at 93 and 145\;GHz. 
 }
\end{abstract}

               

\section{Introduction} \label{sec:intro}

\begin{deluxetable*}{c c c c c c c}

\tablehead{\colhead{Frequency}  & \colhead{FWHM} & \colhead{Baseline} & \colhead{Goal} & \colhead{Frequency} & \colhead{Detector}& \colhead{Number of}\\[-2mm]
\colhead{(GHz)} & \colhead{(arcmin)} & \colhead{($\mu$K-arcmin)} & \colhead{($\mu$K-arcmin)} & \colhead{Bands} & \colhead{Number}& \colhead{SATs}}
\tablecaption{SO SAT Design Specifications\label{tab:SATSensitivity}}

\startdata
    27 & 91& 35 & 25 &\multirow{2}{*}{LF} & 518 & \multirow{2}{*}{1}\\
    39 & 63& 21  & 17 & & 518 &\\
    \hline
    93 & 30 & 2.6 & 1.9 & \multirow{2}{*}{MF}& 12,096 &\multirow{2}{*}{2} \\
    145 & 17 & 3.3 & 2.1 & & 12,096 & \\
    \hline
    225 & 11 & 6.3 & 4.2 & \multirow{2}{*}{UHF}&  6,048 &\multirow{2}{*}{1} \\
    280 & 9 & 16 & 10 & & 6,048 &\\
    \hline
\enddata
\tablecomments{SO SAT projected survey sensitivity with $f_{sky} = 0.1$~\citep{SOGnF2019}. The quoted detector numbers assume seven detector wafers per SAT. LF, MF, and UHF stand for low-frequency, medium-frequency, and ultra-high-frequency respectively. 
}
\vspace{-5mm}
\end{deluxetable*}

Observations of the cosmic microwave background (CMB) are important for developing an understanding of early universe physics. While satellite experiments such as the Wilkinson Microwave Anisotropy Probe (WMAP)~\citep{Bennett2013,Hinshaw2013} and Planck~\citep{PlanckI2018} have produced full-sky microwave maps, ground-based experiments have advanced our understanding by providing higher precision measurements at a variety of angular scales. High-resolution (${\sim}1'$) experiments, such as the Atacama Cosmology Telescope~\citep{Thornton2016, Fowler2007} (ACT) and the South Pole Telescope~\citep{Ruhl2004, Sayre2019} (SPT), have made measurements of both the primordial  power spectrum as well as secondary anisotropies such as the thermal and kinetic Sunyaev-Zeldovich effects (SZE) and gravitational lensing effects.  
Low-resolution experiments (${\sim}0.5^\circ$), such as the BICEP and Keck Array~\citep{BICEP2014, BK2018}, the SPIDER experiment~\citep{Gudmundsson2015}, the Atacama B-mode Search (ABS)~\citep{Kusaka2018}, the POLARBEAR experiment~\citep{POLARBEAR2014, Adachi2019}, the Cosmology Large Angular Scale Surveyor (CLASS)~\citep{Xu2020c}, the Large Scale Polarization Explorer (LSPE)~\citep{Addamo2021}, and the Q and U Bolometric Interferometer for Cosmology (QUBIC)~\citep{Mennella2023} aim to detect B-mode polarization signals at larger angular scales ($\ell\lesssim200$). 

The Simons Observatory~(SO)~\citep{SOGnF2019, Galitzki2018}, located in the Atacama Desert of Chile, is a new CMB experiment consisting of four Small Aperture Telescopes (SAT)~\citep{Galitzki2018, Ali2020, Kiuchi2020}, in the nominal configuration, and one Large Aperture Telescope (LAT)~\citep{Parshley2018, Zhu2021, Gudmundsson2020}. 
Each SAT contains thousands of polarization-sensitive transition edge sensor (TES) detectors~\citep{Duff2016} located on a focal plane cooled to less than 100\,mK and coupled to a refracting optics system and aperture stop held at less than 1\,K. 

Over a five-year observing period, the SATs are projected to achieve a combined map sensitivity of 2\,$\mu$K-arcmin in the combined 93 and 145\,GHz bands at sub-degree angular resolution over $\sim$10\% of the sky~\citep[see Table~\ref{tab:SATSensitivity},][]{SOGnF2019}.  The SAT survey will overlap with the projected BICEP Array coverage~\citep{Hui2018}. The SATs will be able to make precise measurements of the large-scale CMB polarization power spectra in order to detect or constrain the signal from primordial B-modes with a goal sensitivity of $\sigma(r) = 0.002$. A complete discussion of the SAT science objectives can be found in~\cite{SOGnF2019} with a summary of the projected performance shown in Table~\ref{tab:SATSensitivity}. 

In this paper, we start with an overview of the SAT design in Section~\ref{sec:overview} followed by a detailed discussion of the instrument in Section~\ref{sec:design}. The SAT testing and validation results are presented in Section~\ref{sec:int} and Section~\ref{sec:readout}. Finally, we provide a summary of future developments and our results in Section~\ref{sec:conclusion}.

\section{Overview}
\label{sec:overview}

The SATs are optimized to cover the large angular scales ($30 \leq l \leq 200$) containing the predicted peak of the primordial B-mode signal by using a 42\;cm diameter aperture stop coupled to a nearly 40\;cm diameter focal plane. The field of view (FOV) of each telescope is $35^\circ$. 

Each SAT in the array is composed of two primary components. The SAT platform (SATP) provides the mounting and pointing structure for the receiver as well as optical baffling components. The SAT receivers couple light to dichroic detectors with the receivers denoted by one of three frequency band pairings. The frequency pairings are 27/39\;GHz, 93/145\;GHz, and 225/280\;GHz for the SAT-LF (`low-frequency'), SAT-MF (`mid-frequency'), and SAT-UHF (`ultra-high frequency'), respectively. The sensitivity requirements of the experiment necessitate more observing weight on the primary CMB bands so two `mid-frequency' receivers are deployed as the first light instruments, denoted SAT-MF1 and SAT-MF2. The high frequency SAT is designed primarily to measure the Galactic dust emission. The low frequency SAT is designed to measure Galactic synchrotron and free-free emission. The focus of this paper is on the integration and testing of the first mid-frequency SAT, SAT-MF1.

\begin{figure*}
  \centerline{
    \includegraphics[width=1.0\linewidth]
{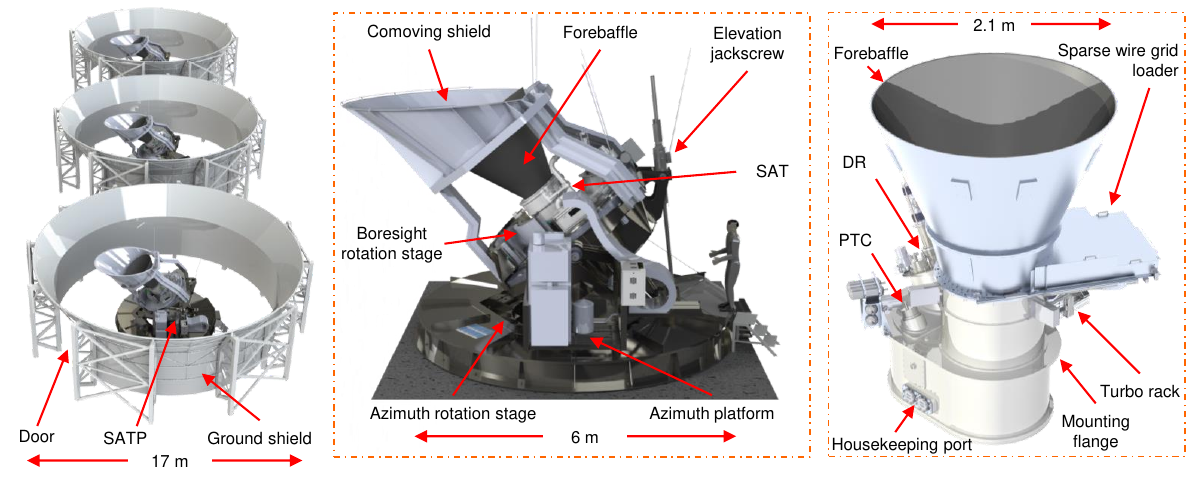}
}
  \caption{ Left: A schematic of an array of three SATPs inside their ground shields. Each SAT receiver is mounted to a SATP and surrounded by a ground shield. Center: Detailed view of the small aperture telescope platform (SATP) with the receiver mounted as reference. Right: Detailed view of the SAT receiver with the forebaffle and sparse wire grid loading mechanism with labelled components including the dilution refrigerator (DR) and pulse tube cooler (PTC).\label{fig:satp}}
\end{figure*}

State-of-the-art millimeter-wave TES detector arrays are photon noise-limited. Thus, increasing detector counts is one of the most straightforward ways to increase experimental sensitivity within a finite observing period. The SATs use dichroic pixels sensitive to both linear polarization directions, each containing four TES detectors~\citep{Duff2016, Duff2024, Choi2018, Stevens2020}. The detectors are packaged into seven hexagonal units per SAT focal plane, referred to as universal focal-plane modules (UFMs).  Each ultra-high and mid-frequency UFM contains 1,720 optical detectors and 36 dark detectors~\citep{Mccarrick2021a, Mccarrick2021b}, giving a total of 12,292 detectors in each ultra-high and mid-frequency SAT. 

A microwave multiplexing system ($\mu$MUX) using SLAC microresonator radio frequency (SMuRF) electronics reads out the detector signals~\citep{Dober2021, Henderson2018, Yu2023}. This multiplexing system is capable of reading out over 1000 detectors via a single coax cable pair, resulting in only two coax pairs per UFM. This greatly simplifies the cryogenic cable management for reading out large format detector arrays. The coax lines are complemented by a woven cable for each UFM that provides the TES bias voltages, the flux ramps for the multiplexing circuit, and the power for the cryogenic low-noise amplifiers (LNAs). The SMuRF electronics systems are contained in two custom liquid-cooled NEMA-4X rated enclosures that are co-mounted with the receiver on the SATP. 

Three 45\;cm diameter silicon lenses with metamaterial anti-reflective coating ~\citep{Golec2020, Coughlin2018, Datta2013} couple the arrays to the sky with diffraction-limited resolution across the focal plane. The aperture stop, all three lenses, and two of the low-pass-edge (LPE) filter elements are combined into a single unit referred to as the optics tube (OT). To maximize the mapping speed, the entire OT is cooled to $<1$\,K--- the first ground-based CMB telescope to do so~\citep{Hill2018}. The OT assembly is 0.26\,m$^3$ in volume and over 200\,kg in mass. 

The SATPs were built by Vertex Antennentechnik GmbH\footnote{Vertex Antennentechnik GmbH, 47198 Duisburg, Germany} in Germany. Each SATP has an azimuthal rotation stage capable of covering up to 540$^\circ$ of rotational freedom and includes an integrated platform for mounting cryogenic equipment and electronics. The elevation stage provides pointing over a range from 20$^\circ$ to 90$^\circ$ above the horizon and is driven by a jackscrew mounted at its rear. The elevation can point to 0$^\circ$, after the removal of part of the optical baffling, for servicing and mounting components. Additionally, each SATP has a boresight pointing stage that allows $\pm 90^\circ$ rotation around the receiver optical axis for assessing possible systematic effects associated with the polarization angle.

\begin{figure*}
  \centerline{
    \includegraphics[width=1.0\linewidth]
{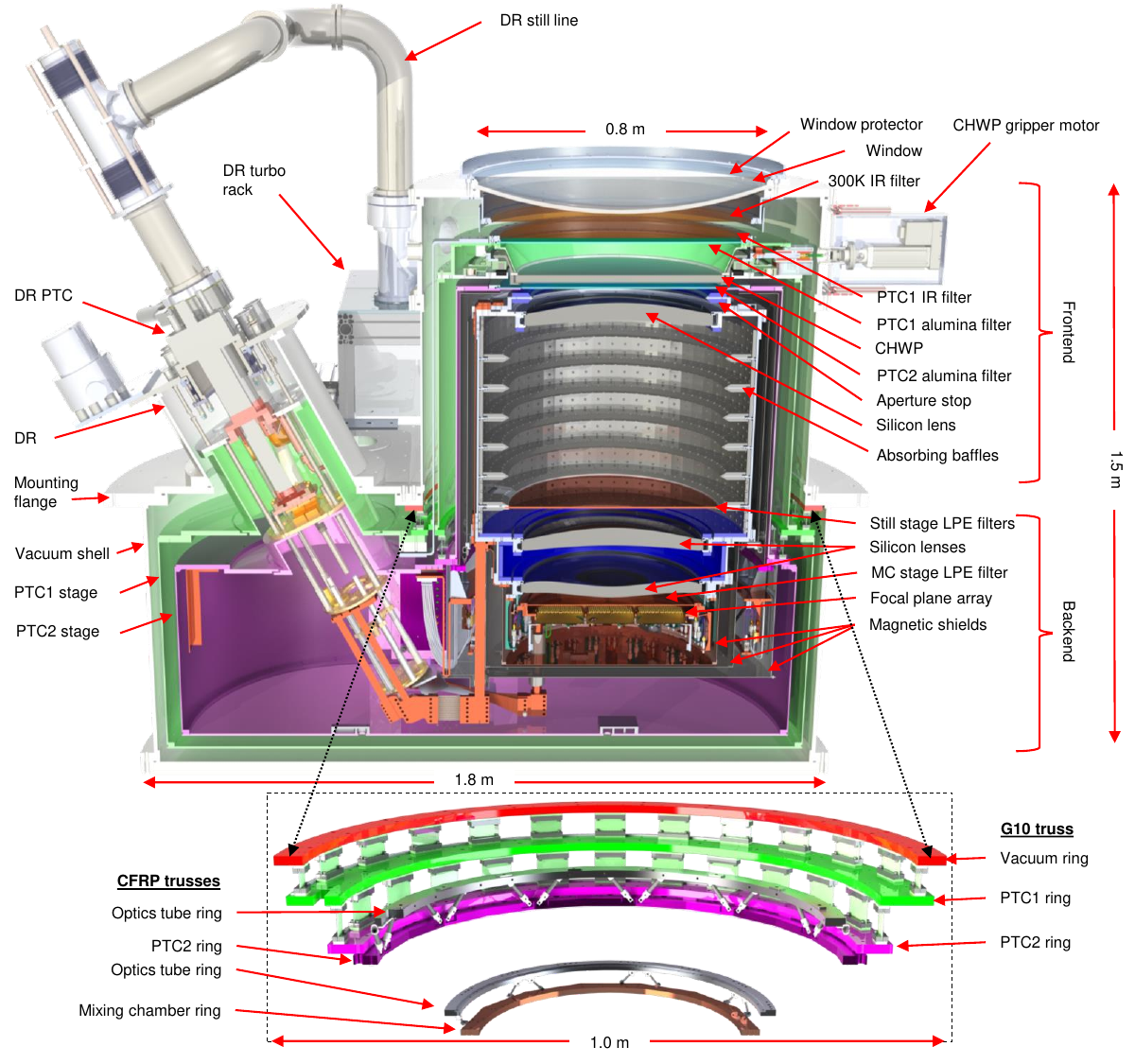}
}
  \caption{ A cross-section view of an SAT showing the principal components. Inset: A detailed view of the three primary G10 fiberglass and carbon fiber reinforced polymer (CFRP) mechanical trusses as they are situated in the overall assembly with all surrounding components hidden.\label{fig:SATX}}
\end{figure*}

Stray light and ground illumination are significant systematics concerns for the SAT. The SAT is designed such that radiation from the ground is diffracted twice before entering the optics at our nominal pointing elevations. This design incorporates a forebaffle mounted to the front of each receiver, a comoving shield attached to the elevation structure of the SATP, and a groundshield that encircles the entire SATP.  Figure~\ref{fig:satp} shows all of these elements in the SAT/SATP design.

The forebaffle is conical with a weatherproofed microwave black coating on the interior surface, an aperture diameter of 2.1\;m, and a height of 1.7\;m from the window surface with the criterion that geometric rays with $>40^\circ$ incidence angle relative to the boresight cannot directly illuminate the receiver window. Additionally, the base of the forebaffle integrates an automated sparse wire grid mechanism that can be inserted in front of the window and rotated to allow for polarization angle calibration during routine observing operations~\citep{Murata2023}. The comoving shield is a conical section with a reflective aluminum surface that is 2.3\;m tall with the tip extending 3.7\;m from the window surface. 

The groundshield is built on a separate foundation surrounding the SATP with the upper edge at a radius of 8.4\;m  and height of 5.66\;m and is formed from cylindrical and conical fiberglass panels. The interior of the cylindrical wall section is covered in reflective zinc-plated steel panels to create a uniform surface facing the SATP. The upper 2\;m of the ground shield forms an aluminum cone angled at 30$^\circ$ from vertical to reflect radiation skyward. At the top of the shield is a 5\;cm radius aluminum pipe composed of eighteen sections and smoothly curved to reduce the diffracted radiation from the ground. The combination of the three shields provides a solution to balance the SAT sensitivity and systematics requirements with project costs and the features of the Chilean site. 

Polarization modulation has become a prominent technique to enable polarization measurements over large angular scales by effectively decoupling the polarization signal from non-modulated signals such as atmospheric effects~\citep{Harrington2021, Simon2016, Hill2020, Reichborn2010, Essinger2016, Takakura2017, Kusaka2014}. The SATs use a cryogenic rotating half-wave plate (CHWP) operating at 50\;K with a superconducting magnetic bearing. The nominal spin speed of the CHWP is 2\;Hz, which modulates the CMB polarized signal at 8\;Hz. The SAT-MF1 CHWP is composed of a three layer sapphire stack sandwiched between two alumina plates each with a two layer mullite/duroid anti-reflective coating \citep{Sugiyama2024, Sakaguri2022}. More details of the design and performance of the CHWP used in the SATs can be found in~\cite{Yamada2023}.

\section{Design}\label{sec:design}

 The scientific goals of the SAT require cooling the detectors, optics, and CHWP system to cryogenic temperatures. This entails tight constraints on the overall design of the SAT-MF1 receiver, including the thermal load on each temperature stage must fit within the capacity of the cooling systems and the system must operate over the rotational range of the telescope. The subsystems and their respective interfaces with the receiver are described below.
 
\begin{figure}
    \includegraphics[width=1.0\linewidth]{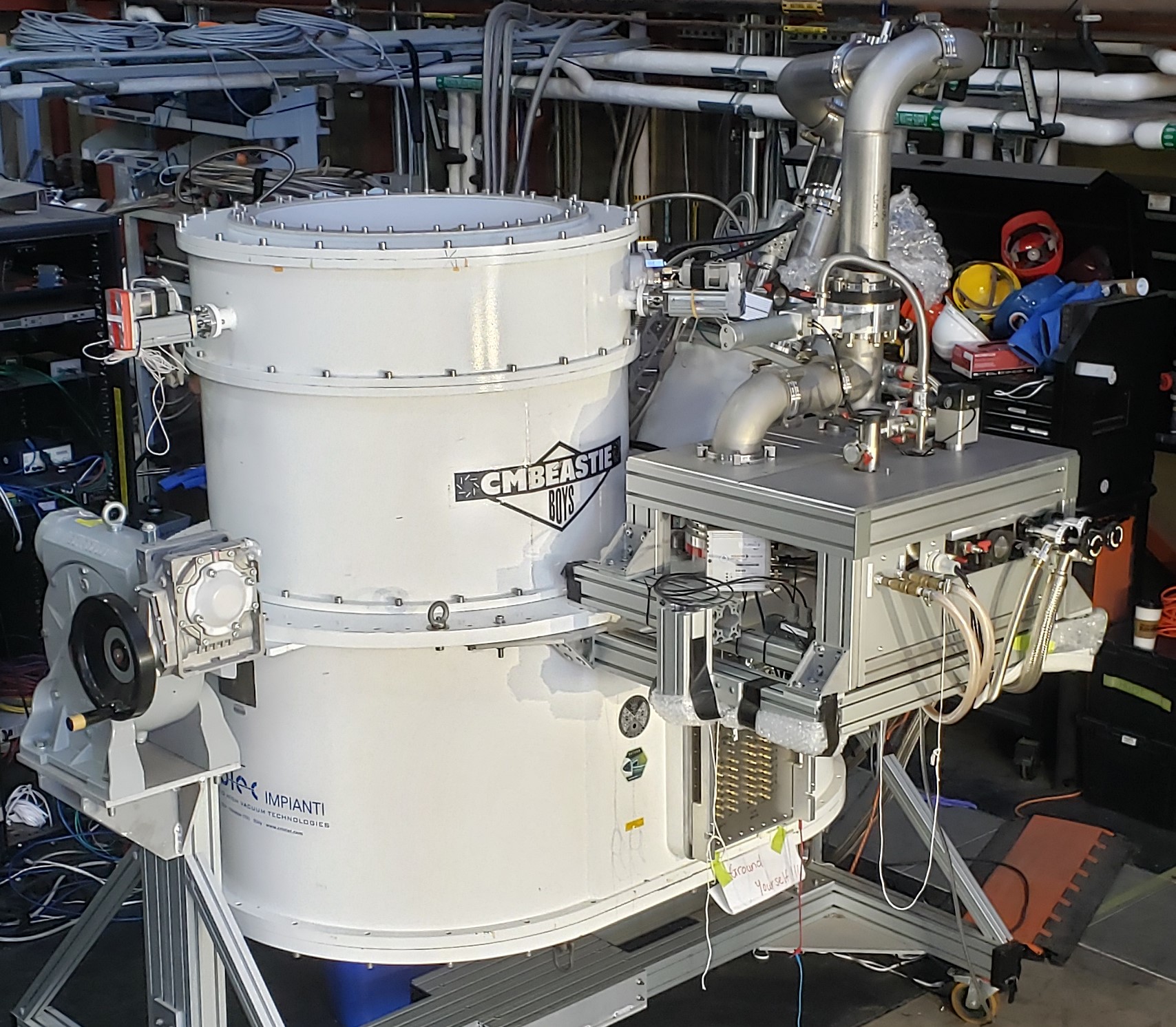}
  \caption{The fully assembled SAT-MF1 in the UC San Diego laboratory space. The rack mounted to the side houses the turbo-molecular pumps for the DR circuit and is attached to the receiver with a mounting jig for testing. 
  \label{fig:SAT_profile}}
\end{figure}

 The SAT structure incorporates an outer vacuum shell with two nested thermal shells cooled by two Cryomech\footnote{Cryomech, Syracuse, NY 13211, USA} PT420-RM pulse tube coolers (PTCs), one of which is integrated into the dilution refrigerator assembly (DR). We refer to the first thermal shell as the PTC1 stage and the second thermal shell as the PTC2 stage. An SD-400 DR manufactured by Bluefors\footnote{Bluefors Oy, 00370 Helsinki, Finland} provides the cooling power for the 1\;K still stage, which is coupled to the OT, and the 100\;mK mixing chamber (MC) stage, which is coupled to the focal plane assembly (FPA). The FPA houses the seven UFMs. The principal elements of the SAT deisgn are shown in Figure~\ref{fig:SATX} and an image of the fully assembled receiver in lab is shown in Figure \ref{fig:SAT_profile}. 
 
\begin{deluxetable*}{l | c c c c c c c | c}

\tablehead{\colhead{Stage} & \colhead{Radiative} & \colhead{Support} & \colhead{Cabling} &  \colhead{LNAs} & \colhead{Optical} & \colhead{Other} & \colhead{Total} & \colhead{Cooling Capacity}\\[-2mm]
\colhead{(K)} & \colhead{(W)} & \colhead{(W)} & \colhead{(W)} & \colhead{(W)} & \colhead{(W)} & \colhead{(W)} & \colhead{(W)} & \colhead{(W)}}

\tablecaption{SAT Thermal Loading Predictions\label{tab:loading}}

\startdata
    PTC1 (45)   & 6.0 & 5.2 & 1.9 & 0.84 & 5.2 & 2.2   & 21 & 50 \\
    PTC2 (4.2)   & 0.002 & 0.17 & 0.14 & 0.056 & 0.095 & 0.12   & 0.58  & 1.8 \\
    Still (1.0)    & $<0.01\times10^{-3}$ & $0.74\times10^{-3}$ & $0.12\times10^{-3}$  & N/A  &  $3.8\times10^{-3}$ & $0.08\times10^{-3}$  & $4.7\times10^{-3}$ & $25\times10^{-3}$ \\
    MC (0.1)  & $<0.01\times10^{-6}$ & $12\times10^{-6}$& $4.2\times10^{-6}$ &  N/A  & $6.8\times10^{-6}$ & $4.7\times10^{-6}$ & $28\times10^{-6}$ & $400\times10^{-6}$ \\
\enddata

\tablecomments{Loading predictions for each temperature stage of an SAT split by source. The cooling power at 45\,K (PTC1 stage) and 4.2\,K (PTC2 stage) is supplied by a PT420-RM PTC. The cooling power at 1\,K (still stage), and 100\,mK (\mc{} stage) is supplied by an SD-400 \dr. The listed cooling capacity is as advertised by the manufacturers and corresponds to the  loading at which the stage will exceed the fiducial temperature. The cooling capacity at 45\;K and 4.2\;K does not include the PT420-RM integrated into the DR, which is coupled to the PTC1 and PTC2 stages. The loading estimates on the PTC stages do not include loading from the \dr{} circulation system. The ``Other'' column includes loading from the aluminized mylar RF shield on all stages as well as the CHWP loading on the PTC1 stage, the array loading on the \mc{} stage, and loading from attenuators on the readout coax chain on the PTC1, PTC2, and still stages. Radiative loading denotes thermal transfer between the shells whereas optical loading is specific to the filter stack. }
\vspace{-5mm}
\end{deluxetable*}

\subsection{Thermal Loading Estimation}
\label{ssec:design-methods}

Thermal loading was estimated on each temperature stage using several standard methods to ensure it sufficiently matched to the cooling capacity of the system. The fiducial values of 45\;K, 4.2\;K, 1.0\;K, and 0.1\;K are used for the PTC1, PTC2, still, and MC stages, respectively, to simplify the calculation of the thermal loading on each stage. Radiative coupling between stages, excluding the optical path, was estimated using the Stefan-Boltzmann law. The emissivity of each surface was estimated from the material properties of aluminum and an effective emissivity based on the 40 layer multi-layered-insulation (MLI) blankets from Ruag\footnote{Ruag Space GmbH, 1120 Vienna, Austria} on the PTC1 stage and the 10 layer blankets on the PTC2 stage. The radiative loading at the still and MC stages from the PTC2 stage thermal emission is negligible and so no MLI is employed at those stages.

The conductive loading is derived from a standard integrated conductivity across the material, 
\begin{equation}
    P_{thermal} = \frac{A}{L}\int_{T_{low}}^{T^{high}}\kappa(T)dT,
    \label{eq:therm_cond}
\end{equation}
where $A$ is the cross-sectional area of the component with length $L$ and temperature dependent conductivity $\kappa(T)$. Conductivity data from~\cite{Marquardt2002, Woodcraft2009, Woodcraft2010, Crowley2022} were used. The conductivity calculations are employed for the mechanical supports as well as any electronic cabling. Electrical loading from other readout component such as LNAs or attenuators are based on measurements or manufacture specifications.

The thermal loading from the optical path at each stage was predicted using a ray tracing solver to estimate the radiative transfer between elements, which was also used to make estimates for the SO LAT receiver \citep{Zhu2018, Zhu2021}. The program takes as inputs the dimensions of the system, the emissivity of wall materials, and the transmission, reflection, and material conductivity of the filters. The program produces 100,000 rays at random locations with Lambertian emission directions on each optical surface for each of seven logarithmically spaced frequency bands between 0 and 10 PHz ($\sim$30\;nm, ultraviolet). The filter spectral data and surface emissivity values determine the probability a ray is reflected, absorbed, or transmitted with the number of surface interactions limited to 1000. Filter data do not typically cover the entire frequency regime simulated as filter measurements are focused on the frequency regime around the anticipated blackbody radiation peak for a given filter. Data are projected beyond the measured spectrum to higher and lower frequency regimes by setting a constant value equal to the last measured data point. The absorbed fraction is used to calculate Gebhart factors~\citep{Gebhart1961},   

\begin{equation}
    B_{ij} = \frac{\Delta P_{ij}}{\epsilon_i A_i \int_{\nu_1}^{\nu_2}B_\nu (T_i)d\nu}\,,
    \label{eq:gebhart}
\end{equation}
\noindent where $\Delta P_{ij}$ is the power absorbed at surface $A_j$ emitted by surface $A_i$ with emissivity $\epsilon_i$. $B_\nu(T)$ is the Planck function that describes the surface brightness integrated over solid angle per unit frequency interval. The Gebhart factors are useful as an analytic tool to understand how radiation is coupled between surfaces. From Equation~\ref{eq:gebhart} we can get an expression to calculate the net power transferred between two surfaces $i$ and $j$,
\begin{equation}
    \Delta P_{ij,net} = \epsilon_i A_i B_{ij} \int_{\nu_1}^{\nu_2}\left[ B_\nu (T_i)-B_\nu(T_j)\right]d\nu\,.
    \label{eq:radtran}
\end{equation}

The temperature of wall surfaces is fixed at the input values while the filters are divided into 10 equal-width annuli with thermal equilibrium achieved through non-linear minimization of total conductive and radiative power. The output predictions of the loading on each stage are used to estimate the loading along the optical path of teh SAT with elements described in Section \ref{ssec:design-opt} and values summarized in Table \ref{tab:loading}. 

\begin{figure}
    \includegraphics[width=1.0\linewidth]{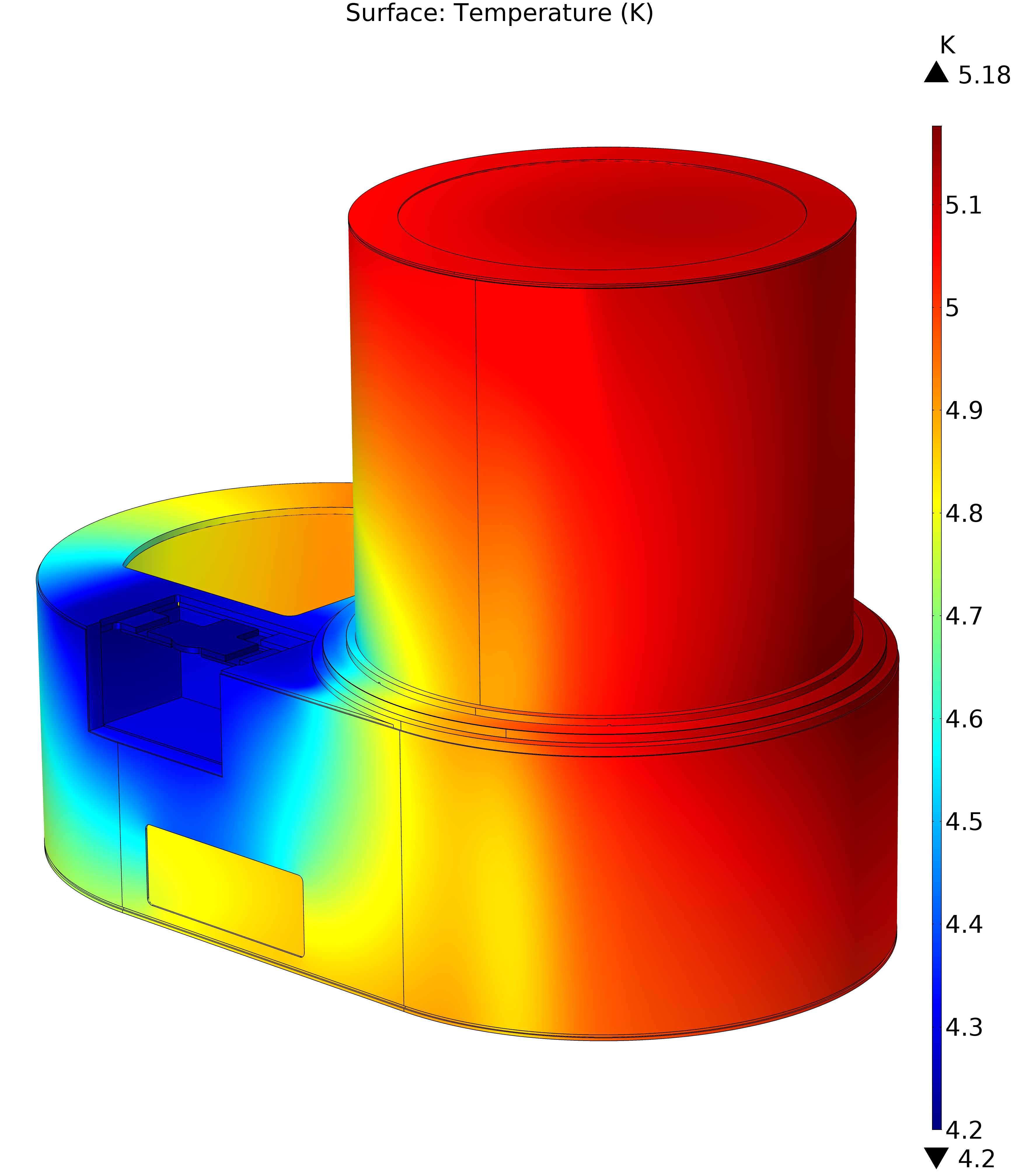}
  \caption{Example of a COMSOL simulation to assess thermal gradients across the PTC2 stage with the fiducial coldhead value of 4.2\;K used as a conservative upper bound. The stage walls use 1100 aluminum, the stage plates use 6061 aluminum, heat straps are composed of a combination of 4N aluminum and OFHC copper, and all surface contacts are assumed to be ideal. Though the actual Al heat straps are 5N8, 4N Al serves as a reasonable approximation and leads to a more conservative estimate of the thermal gradient.  
  In the design phase the simulations guided materials choices and heatstrap locations. In the testing phase, we used the simulations to assess measured temperatures and gradients, and set upper bounds on realized loading.
  \label{fig:comsol}}
\end{figure}

We also performed simulations using COMSOL multiphysics\footnote{COMSOL, Inc., Burlington, MA 01803} to estimate the temperature gradients across each thermal stage of the SAT. These gradient predictions are particularly important in assessing our assumptions of fiducial temperature values used in loading calculations. The simulations also informed our design by highlighting whether we had sufficient thermal paths between load sources and coldheads and indicating what level of cooling capacity overhead was needed to allow cold heads to reach lower temperatures than their fiducial values to compensate for gradients. For instance, the requirement that the UFMs operate at 100\;mK necessitates that the MC at the DR actually be cooler than 100\,mK, with an associated drop in cooling capacity. This thermal gradient informed the choice of heat-strap cross section, allowing sufficient cooling capacity overhead at the DR.

The geometry of each stage was imported to COMSOL from Solidworks\footnote{Dassault Syst\'emes, SolidWorks Corporation, Waltham, MA 02451} with elements such as bolt patterns or cabling that do not affect the thermal properties removed. Values from the thermal calculations in Table \ref{tab:loading} were input as fixed loading values on the respective surfaces with the coldhead temperatures fixed at the fiducial values. Radiative loading was distributed across coupled surfaces while localized component loading was applied to their interface surfaces. We did not include thermal resistance across component interfaces, instead using the default setting which assumes the material thermal conductivity. We also did not include the effect of gold plating on components where present. The effect from thermal resistance at interfaces is assumed to be sub-dominant to gradients from the material conductivity, but their exclusion means the simulation results serve as a useful lower limit on the thermal gradient. An example of the analysis can be seen in Figure \ref{fig:comsol}. We also performed simulation sets with non-fiducial coldhead temperatures using load curves, including PTC load curves at different tilt angles relative to gravity, to more fully explore the predicted temperature gradients.

\subsection{Cryogenic Architecture}
\label{ssec:design-cryo}

The thermal model of the SAT receiver derived from the subcomponent properties and summarized in Table \ref{tab:loading} sets the requirements on the cooling system and thermal conduits. 
A Cryomech PT420-RM provides the primary heat lift at the PTC1 and PTC2 stages and a modified SD-400 Bluefors DR provides the cooling power for the still and MC stages, with  cooling capacities shown in Table \ref{tab:loading}. The DR also incorporates a PT420-RM unit to precool the circulated helium and to provide additional cooling capacity to the PTC stages. 

 The operational elevation and boresight pointing ranges are set by the requirement that the cryogenics not tilt more than $45^\circ$ away from vertical as the cooling capacity of the PT420-RM units drops out of an acceptable regime beyond that angle~\citep{Tsan2021}. The cryogenic systems are mounted at an angle of $27.5^\circ$ with respect to the boresight axis to allow a larger elevation pointing range for the telescope with the trade-off of a smaller boresight rotation range at low pointing elevations.  By tilting the cryogenics the telescope is able to point from $90^\circ$ to $17.5^\circ$ above the horizon at a boresight rotation of $0^\circ$ and cover a boresight rotation range of $\pm 75^\circ$ at the nominal scanning elevation of $50^\circ$ above the horizon.

\begin{figure}
    \includegraphics[width=1.0\linewidth]{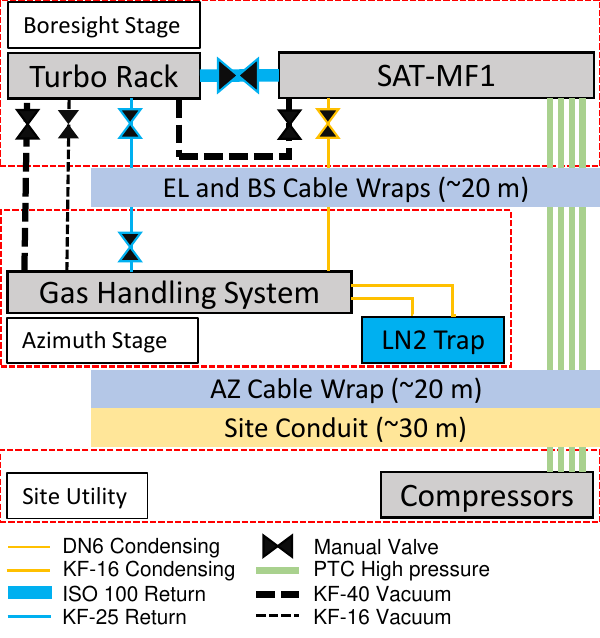}
  \caption{ Schematic diagram of the helium and vacuum systems on the SAT including all hoses between the PTC compressors, turbo rack, GHS, and the SAT itself with manual valves called out. The location and length of the SATP cable wraps for azimuth (AZ), elevation (EL), and boresight (BS) as well as the conduit run are identified. The vacuum lines are installed temporarily for service as they are not used in operation and thus do not route through the cable wrap. The KF-40 line pulls vacuum from the SAT while the KF-16 line simplifies servicing and pulling vacuum on he DR circuit. Red dashed lines denote distinct zones of the telescope where equipment is mounted.
  \label{fig:cryoschem}}
\end{figure}

A variety of heat strap solutions were implemented across the PTC and DR stages with an emphasis on the conduits from the DR coldheads as the OT and the FPA have the most stringent temperature requirements of the system. The entire OT assembly must be at $<1$\,K and the FPA must be at $<100$\,mK.  The still and MC cold heads of the DR are coupled to their respective stages through a combination of oxygen-free high conductivity (OFHC) copper rods welded to mounting base plates and flexible heat straps provided by TAI\footnote{Technology Applications, Inc., Boulder, CO 80301} that create a mechanical break between the stages and the cold heads. All principal components are gold plated to improve interfaces and conductivity. The flexible straps couple to the rods via clamping blocks with geometry informed by~\cite{Didschuns2004}. Custom Invar washers were installed along with Belleville disc spring washers at heat strap clamp interfaces with stainless steel bolts going through copper for the still and MC stages. The thickness of the Invar was set such that the bolt and washer assembly would contract the same amount as the copper at cryogenic temperatures in order to maintain the clamping force. The still heat strap assembly is stiffened using four 3\;mm diameter carbon fiber reinforced polymer (CFRP) tube stand-offs from the MC stage. The MC flexible heat-strap is supported along much of its distance by an aluminum channel assembly to allow us to control the free length and stiffness of the joint. 

The cryogenic system of the DR has several modifications to facilitate its use in a pointed telescope platform. Additional gas gap heat switches were added by Bluefors between the PTC2, still, and MC stages for four total between each stage to reduce cool down times. The pulse tube was upgraded from a PT410-RM to a PT420-RM to provide additional cooling power at each stage. The SAT receiver shells and the DR PTC stages are coupled via 0.999998 (5N8) purity aluminum foil straps that act as a flexible link between the DR and SAT stages to provide additional cooling overhead to the SAT PTC stages. Additionally, an integrated cold trap coupled to the DR PTC1 stage was added to the DR to complement the external liquid nitrogen cooled trap and function as a safety feature in case the LN2 trap runs out due to site access issues during the anticipated year-long continuous operation of the SATs between service intervals. 

The warm components of the DR also have a number of modifications to improve functionality on the pointed platform. The gas handling system (GHS) is mounted in a custom weatherproof enclosure on the azimuth stage of the SATP. A separate weatherproof turbo rack houses the two Pfeiffer\footnote{Pfeiffer Vacuum Gmbh, Asslar, Germany, 35614} Hi-Pace 400 pumps that are the primary source of He mixture circulation. 

The turbo rack is mounted next to the SAT receiver on the boresight stage of the SATP. A 20\;m long, 6\;mm diameter flexible line routes the helium from the GHS to the condensing side of the DR circuit.  The helium exits the DR unit via an ISO-100 pipe that carries it to the turbo rack via a Bluefors-provided vibration isolation tee which allows mechanical flexibility between the mating components. A Swagelok valve at the entrance to the DR unit on the condensing line and a gate valve at the exit of the DR unit on the ISO-100 pipe allow isolation of the DR unit during integration operations that require disconnecting from the GHS. A 20\;m KF-25 return line connects the turbo rack to the GHS and routes through the boresight and elevation cable wraps along with the condensing line. See Figure \ref{fig:cryoschem} for more details of the helium circulation system. The turbo pumps are liquid cooled via a coolant control system on the boresight stage fed by hoses routed through the cable wraps to a chiller that is adjacent to one of the utility containers located external to the SAT groundshield (see Figure \ref{fig:satp}).  

\subsection{Mechanical Design}
\label{ssec:design-mech}

\begin{deluxetable}{c c c c}

\tablehead{\colhead{\multirow{2}{*}{Stage}} & \colhead{Width} & \colhead{Length} & \colhead{Number of} \\[-3mm]
\colhead{} & \colhead{(mm)} & \colhead{(mm)} & \colhead{Tabs} }
\tablecolumns{4}

\tablecaption{G10 Truss Specifications\label{tab:g10_tab}}

\startdata
	Vacuum to PTC1  & 55 & 39.5 & 24\\
	PTC1 to PTC2 & 55 & 54 & 24 \\
\enddata

\tablecomments{Length refers to the thin profile web component between the thicker top and bottom bases of the monolithic G10 piece. All webs are 1.0\,mm thick with a 1.0\,mm radius fillet between the web and the bases. The bases are 5mm thick (see Figure \ref{fig:g10truss}).
}
\end{deluxetable}

\begin{figure}
  \centerline{
    \includegraphics[width=1.0\linewidth]
{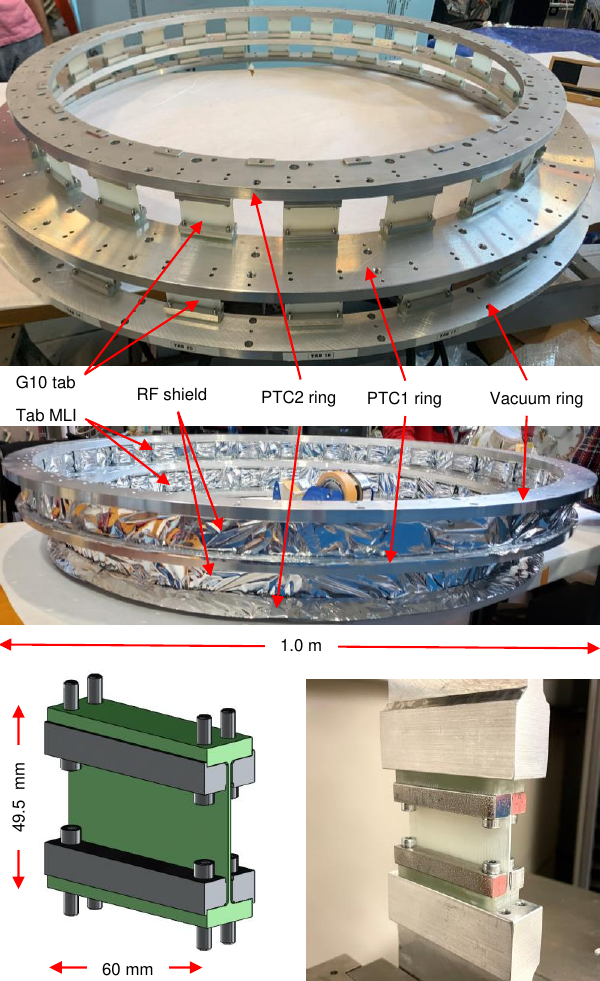}
}
  \caption{ Top: The SAT-MF1 G10 vacuum/PTC1/PTC2 mechanical truss (inverted). Middle: The assembled truss after MLI has been attached to individual tabs and the RF shields have been installed but before the MLI skirts have been added. There are 40 layers of MLI on the PTC1 stage and 10 layers on the PTC2 stage. See Figure \ref{fig:SATX} for more details of the truss design. left: CAD image of a single vacuum to PTC1 G10 monolithic truss tab. Bottom right: The testing setup of a vacuum to PTC1 tab on the Instron universal testing machine. A series of tabs were tested to destruction as part of our development program and subsequent validation of the SAT G10 truss. \label{fig:g10truss}}
\end{figure}

The mechanical viability of components including the vacuum shell and support elements was determined through finite element analysis (FEA) using SolidWorks simulation. The FEA was used to determine the factor of safety (FoS) of the various mechanical components as well as to predict the relative physical and angular displacement of elements over the elevation pointing range of the telescope which could affect the optical performance. Additional vibrational mode analysis was performed on elements attached to the still and MC stages to identify resonance modes that could induce heating during operations. This was used to inform design changes to increase rigidity where needed.

The vacuum shell holds sea-level atmospheric pressure with a minimum FoS of four. The overall mass was not a strong driver of the design. The back-end is closed off with one large lid which allows ample access space to the back-end cryogenics, focal plane, and electronics. This greatly simplifies the integration of components and allows for quick turn-arounds between cool-downs. A flange on the back-end vacuum shell provides the mating surface to the SATP.

The SAT receiver's internal support structure consists of three distinct trusses that provide mechanical support while thermally isolating stages at different temperatures. Each truss mounts near the center of mass of the components it supports to minimize any deflection at non-vertical pointings. The design has a single mating surface at each stage. With the mounting interface near the center of mass, the receiver is naturally split into two primary volumes. The front-end volume is cylindrical and encompasses the majority of the optical components. The back-end volume is oval in shape with flat sides to accommodate access ports and electronics feed-throughs and encompasses the cryogenic components, focal plane, and cold electronics components. Details of the design can be seen in Figure \ref{fig:SATX} and Tables \ref{tab:g10_tab} and \ref{tab:CF_strut}.

The first support truss (the G10 vacuum/PTC1/PTC2 truss) attaches to the same plate that mates to the SATP. The truss is composed of three rings corresponding to the vacuum shell, PTC1 stage, and PTC2 stage with the thermal offset between stages provided by G10 tabs that bolt to the rings. Details of the G10 vacuum/PTC1/PTC2 truss design are shown in Figure \ref{fig:g10truss} and Table \ref{tab:g10_tab}. 

The second truss is composed of CFRP struts running between two aluminum rings at $\sim$45$^{\circ}$ and provides the mechanical support for the still stage, with one end bolted to the PTC2 ring of the G10 vacuum/PTC1/PTC2 truss and the other end bolted to a flange near the center-of-mass of the OT (referred to as the CFRP PTC2/OT truss).    The struts providing the mechanical support and thermal isolation between the rings are comprised of CFRP tubes with aluminum caps epoxied with 3M Scotch-Weld 2216\footnote{3M Company, 2501 Hudson Rd, Maplewood, MN 55144, USA} on each end of the tubes which are then bolted to the aluminum rings. The CFRP tubes are manufactured by Clearwater Composites\footnote{Clearwater Composites, LLC., 4429 Venture Avenue, Duluth, MN 55811, USA}. The CFRP PTC2/OT truss is re-entrant with respect to the G10 vacuum/PTC1/PTC2 truss to keep the mounting points as close to the center-of-mass of the system as possible. 

A third, smaller CFRP truss (the CFRP OT/FPA truss) of similar design connects the focal plane on the MC stage to the bottom surface of the OT. The CFRP tubes for this truss are manufactured by vDijk Pultrusion Products\footnote{vDijk Pultrusion Products, Aphroditestraat 24, NL-5047 TW TILBURG, The Netherlands} (DPP). More details of the design, manufacture, and testing of the CFRP struts is described in~\cite{Crowley2022}.

A sheet of 6.25\,\micron thick mylar with an $\approx150$\,\AA{} aluminum layer on both sides, procured from MEI\footnote{MEI - Metallized Engineering, Suffield, CT 06078}, bridges each thermal gap on all the support trusses to create an enclosed Faraday cage encompassing the back-end.  The back-end vacuum shell provides the primary element of the RF shield with specialized compressible RF gaskets\footnote{Parker Chomerics, Woburn, MA 01888} installed on the outside of each vacuum O-ring to ensure the shielding is maintained at vacuum joints.  Spectrum Control\footnote{Spectrum Control, 8061 Avonia Road
Fairview, PA 16415, USA} Pi filters are attached at the hermetic sockets with type 56-745-005 and type 56-745-003 for the housekeeping and non-coax readout inputs, respectively. The alumnized mylar then carries the shield down to the focal plane where the FPA itself completes the shield in order to isolate electronic components in the back-end volume from RF-interference. 

\begin{figure}
    \includegraphics[width=1.0\linewidth]{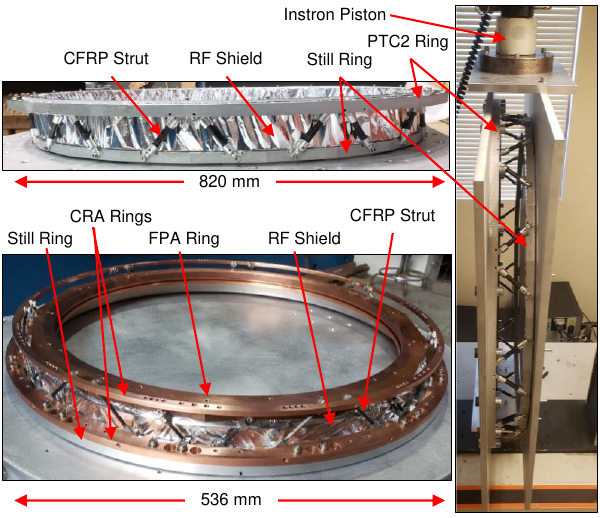}
  \caption{Top Left: The CFRP PTC2/OT truss with the RF shield installed. Bottom Left: The CFRP OT/FPA truss with the RF shield installed. The carbon fiber supports and copper ring for the CRA components at the MC stage is also integrated into the pictured assembly.  Right: Validation of the CFRP PTC2/OT truss in the Instron universal testing machine for the shear force test.
  \label{fig:CFtruss}}
\end{figure}

The design also incorporates two additional CFRP trusses for the cold readout electronics from the PTC2 to OT stages and from the OT to FPA stages. More details of the readout architecture are described in Section \ref{ssec:design-readout}. The mass suspended by these trusses is only a few kilograms and their primary purpose is to provide a modular mechanical structure for the readout components that is stiff, with primary resonance modes above $\sim40$\;Hz, and low thermal conductance. Details of the CFRP trusses are shown in Figures \ref{fig:SATX} and \ref{fig:CFtruss} and Table \ref{tab:CF_strut}. 

\subsection{Truss validation}
\label{ssec:int-mech}

The SAT trusses had to meet two primary mechanical goals: 1) support the system with enough margin of safety to ensure reliable operation over all observing, transport, and installation procedures, and 2) maintain the alignment of the optical system to within the tolerances specified by the optical design (see Section \ref{sssec:int-g10align}). 

The initial G10 truss tab design did not meet the mechanical requirements of the receiver which necessitated a second design iteration. Both the initial design and testing sequence as well as the subsequent redesign and validation process are described in this section. 

\subsubsection{G10 vacuum/PTC1/PTC2 truss strength tests}
\label{sssec:int-g10strength}

The G10 vacuum/PTC1/PTC2 truss requirements are determined from the most stringent use case of the assembly in three primary configurations: axial pull, axial compression, and shear (see Tables \ref{tab:truss_req} and \ref{tab:strut_req}). During shipping, six helical isolators are mounted between the platform the SAT is mounted to and the base of the crate that surrounds the SAT to provide the shock and vibration protection. The transportation of a fully assembled SAT across the SO site to the SATP is performed on the same shock isolated platform.  The platform reduces a 10\,g external shock to less than 3\,g along any axis at the receiver. While 10\,g shocks are typically seen during shipping, the site transport procedure is designed to not impart any significant shock loads, thus the 3\,g requirement serves as a conservative upper-bound and is the most stringent requirement on the truss. The transport procedure was executed without issue in 2023. 

The first version of the G10 vacuum/PTC1/PTC2 truss was manufactured with tabs composed of a G10 tab epoxied between two aluminum feet that mounted to the truss rings. Results of pull tests to destruction of witness samples on an Instron universal testing machine\footnote{Instron, Norwood, MA 02062, USA} found a typical 0.05\% offset yield strength of $>2.0$\;kN, exceeding our initial requirements. The full trusses were tested with partial liquid nitrogen submersion to simulate differential thermal contraction, followed by hang tests in shear to approximately half the expected supported mass. The initial validation program insufficiently probed the effect of radial contraction of the PTC1 ring relative to the vacuum ring that ultimately caused a failure of the truss installed in SAT-MF1 after 9 complete thermal cycles. Fortunately, minimal damage occurred to other components.

A detailed analysis determined that the stresses induced on the epoxy joints at the base of the tabs from the differential thermal contraction had been significantly underestimated in our simulations. The tabs have a relatively low aspect ratio of height--to--width and the contraction forces the tabs into an S-bend that imparts higher stresses at the base of the tab at the epoxy joint.  Additionally, the circular truss is mounted to a vacuum plate on which the deformation due to atmospheric pressure is asymmetric (see Figure \ref{fig:SATX}). Tabs on either side of the deflection experience additional loading  and the stress is concentrated at the corner of the epoxy joint of those tabs. 

The two effects combined caused a failure of the epoxy joint at the tabs at the edge of the cryostat which then caused a gradual cascading failure of the rest of the truss.  It is not clear if additional validation steps would have identified the issue as the LN2 dunk did not replicate the vacuum plate deflection which is difficult to produce outside of the cryostat. A tiger team review, which included outside experts, used both Solidworks and NASTRAN\footnote{\url{https://software.nasa.gov/software/LAR-16804-GS}} FEA to improve our understanding of the forces and moments acting on the tabs. The simulations were able to recreate the observed failure mode. Two other SATs had begun to experience this issue after a smaller number of thermal cycles, which allowed us to localize which tabs failed first and helped confirm this failure mode identified in simulations.

A monolithic G10 I-beam tab machined out of a single piece that eliminates all epoxy joints was determined by simulations and prototype testing to meet our requirements, as shown in Figure \ref{fig:g10truss}. The monolithic design makes the manufacture more repeatable while also removing significant uncertainty from the simulations. Machining the tab also allowed us to control the fillet dimensions where the vertical tab transitions into the base which was critical to increase the strength of the tab to distribute the stresses experienced during vacuum and thermal cycling. To provide resilient mounting of the tab, a steel clamping block is placed between the screw heads and the base of the G10 to more evenly distribute the clamping force on the tabs. The design maintained an identical interface to the existing truss rings. 

\begin{deluxetable}{c c c c c c}

\tablehead{\colhead{\multirow{2}{*}{Truss}} & \colhead{\multirow{2}{*}{Material}} & \colhead{OD} & \colhead{ID} & \colhead{Length} & \colhead{Quantity} \\[-3mm]
\colhead{} & \colhead{} & \colhead{(mm)} & \colhead{(mm)} & \colhead{(mm)} & }
\tablecolumns{4}

\tablecaption{Carbon Fiber Truss Specifications\label{tab:CF_strut}}

\startdata
	Primary: \\
        \hline
        PTC2 to OT  & Clearwater & 8 & 7 & 31 & 24\\
        OT to FPA & DPP & 3 & 2 & 19 & 14 \\
        \hline
        Readout:\\
        \hline
	PTC2 to OT & DPP & 8 & 7 & 141 & 14 \\	
	OT to FPA & DPP & 3 & 2 & 32 & 14 \\
\enddata

\tablecomments{The geometry and number of struts. Length refers to the free length of CF between the end caps. All truss elements are oriented at $\sim$45$^{\circ}$. `Primary' trusses are the two mechanical support CFRP trusses described in detail in Section \ref{ssec:design-mech}. `Readout' trusses are the additional two trusses supporting readout components described in Section \ref{ssec:design-readout}. 
}
\vspace{-5mm}
\end{deluxetable}

\begin{figure}
\centering
  \includegraphics[width=1.0\linewidth]{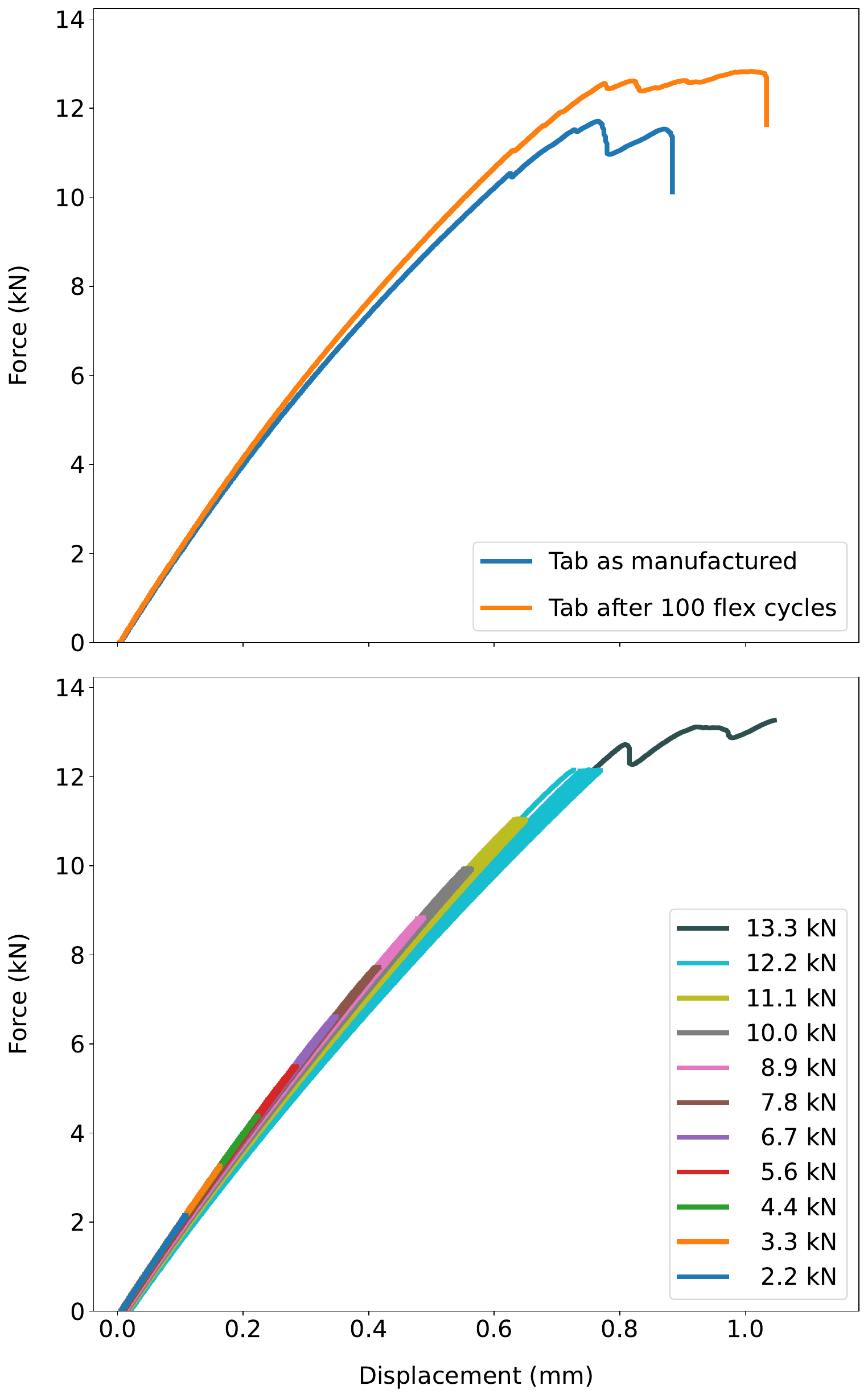}
  \caption{Monolithic G10 tabs were tested to destruction to ensure the design met our specifications (see Table \ref{tab:g10_tab}). Top: The results from testing two vacuum to PTC1 type tabs. The blue line is from a tab as manufactured. The orange line shows a tab that, prior to being tested, has been flexed 100 times with the two ends of the tab displaced horizontally by 2\,mm to simulate the maximum expected flex from the differential contraction of the truss rings.  Bottom: Fatigue testing of a tab which was repeatedly stressed to different loading set points 10 times at each set point before its ultimate failure. All three tabs exceeded our requirements with typical ultimate strengths $>$10.5\,kN \label{fig:g10tabs}}
\end{figure}
 
Monolithic tabs were tested to destruction in the pull direction under three conditions: as manufactured; after mechanically offsetting the top and bottom of the tab by 2\,mm in a CNC mill jig 100 times to simulate the stress from thermal contraction of the truss rings; and after repeated loading in incremented steps with 10 cycles at each set point to probe fatigue effects (see Figure \ref{fig:g10tabs}). All three testing conditions yielded an ultimate strength $>$10.5\,kN, with the failure mode and level consistent with simulations. The tests established that the new tabs would exceed performance requirements in all simulated stress conditions. For the straight pull requirement of 0.575\,kN, the tabs have a FoS of $>$18 (see Table \ref{tab:strut_req}).  

A completed second version truss was installed and tested in an empty SAT vacuum shell with a rigid plate installed onto either the PTC1 or PTC2 truss ring. A load cell was attached to the plate, with a threaded rod attached to the other side of the load cell, and the other end of the threaded rod attached to an aluminum beam mounted across the base of the vacuum shell. Turning the threaded rod adjusted the loading on the truss with the rod mounted at the location of the center-of-mass of the respective stage. The PTC1 and PTC2 stages were loaded to $>3$\,g in tension with forces guided by the values in Table \ref{tab:truss_req}. The maximum load was 15\,kN on the PTC1 ring of the truss with coordinate measuring machine (CMM) measurements before and after the test, combined with visual inspection, showing the truss maintained the required flatness and alignment tolerances. After the new trusses were validated by both the extensive simulation work that was done in combination with the tab testing and full truss testing, the new truss was installed in SAT-MF1 and the receiver testing program was resumed. There has been no indication of further issues with an inspection performed after multiple subsequent cooldowns.

\begin{deluxetable}{l | c c c c}

\tablehead{\colhead{Truss} & \colhead{Mass} & \colhead{Tension} & \colhead{Compression} & \colhead{Shear} 
}
\tablecolumns{5}

\tablecaption{Truss Loading Requirements in g\label{tab:truss_req}}

\startdata
    G10 Vac/PTC1 & 460 kg & 3 & 1 & 1 \\ 
    G10 PTC1/PTC2 & 300 kg & 3 & 1 & 1 \\
    CFRP PTC2/OT & 210 kg & 1 & 3 & 1 \\
    CFRP OT/FPA & 30 kg & 3 & 1 & 1 \\
\enddata

\tablecomments{ The mass supported by each truss in the fully assembled SAT is indicated. `Tension', `Compression', and `Shear' indicate the maximum loading condition from transport or operation along each axis, in multiples of the supported mass.
}
\vspace{-5mm}
\end{deluxetable}

\begin{deluxetable}{l | c c}

\tablehead{\colhead{Strut Type} & \colhead{Tension Requirement} & \colhead{Measured Yield} 
}
\tablecolumns{3}

\tablecaption{Tab and Strut Testing Results\label{tab:strut_req}}

\startdata
     Monolithic G10 tab & 0.575\,kN & $>10.5$\,kN\\ 
     3mm OD CFRP & 0.3\,kN & $>0.8$\,kN \\
     8mm OD CFRP & 1.5\,kN & $>2.5$\,kN \\
\enddata

\tablecomments{ The elastic strength tension requirement for an individual element was calculated using simulations of the full truss loads shown in Table \ref{tab:truss_req}. The measured value refers to the minimum top of the elastic regime for all witness sample tabs and struts which were tested to destruction. More details on the geometry of the struts can be found in Tables \ref{tab:g10_tab} and \ref{tab:CF_strut}.
}
\vspace{-5mm}
\end{deluxetable}

\subsubsection{G10 vacuum/PTC1/PTC2 truss alignment tests}
\label{sssec:int-g10align}

The G10 vacuum/PTC1/PTC2 truss of the SAT provides the alignment of the OT with the frontend optics comprised of the filters, CHWP, and window as well as external baffling. Because the OT contains all the lenses and the FPA coupling, the tolerance on the G10 vacuum/PTC1/PTC2 truss is relatively relaxed which simplifies the manufacture and validation processes. The second truss 
was assembled using a CMM  arm to ensure that, not only were the rings aligned as required, but also that the tabs were consistently mounted to avoid introducing new sources of stress that would not have been captured in simulations. 

The principle requirements on the G10 vacuum/PTC1/PTC2 truss were: tip/tilt of $<0.03^\circ$ between any of the truss rings, distance between any of rings along the optical axis to within $<0.4$\;mm of the specified distance, and concentricity of the rings of $<1$\;mm. CMM metrology was used to confirm the assembled truss met these requirements.

\subsubsection{CFRP truss strength tests}
\label{sssec:int-cfrpstrength}

The loading requirements for the two CFRP trusses are shown in Table \ref{tab:truss_req}. The multiple of 3 on the requirements for the PTC2/OT truss in compression and the OT/FPA truss in tension comes from the site transport condition of the fully loaded SAT receiver described in Section \ref{sssec:int-g10strength}. The other requirements with a multiple of 1 are set by supporting the mass during normal operation.

The most stringent CFRP requirements for both trusses come from the axial loading in the site transport case, from the assembly highbay to the observing platform, with the multiple of three included in the numbers shown in Table \ref{tab:strut_req}. The resulting strut design, analysis, and testing is described in detail in~\cite{Crowley2022}, and the results described therein led to the struts and trusses presented in this paper. 

\begin{figure}
    \includegraphics[width=1.0\linewidth]{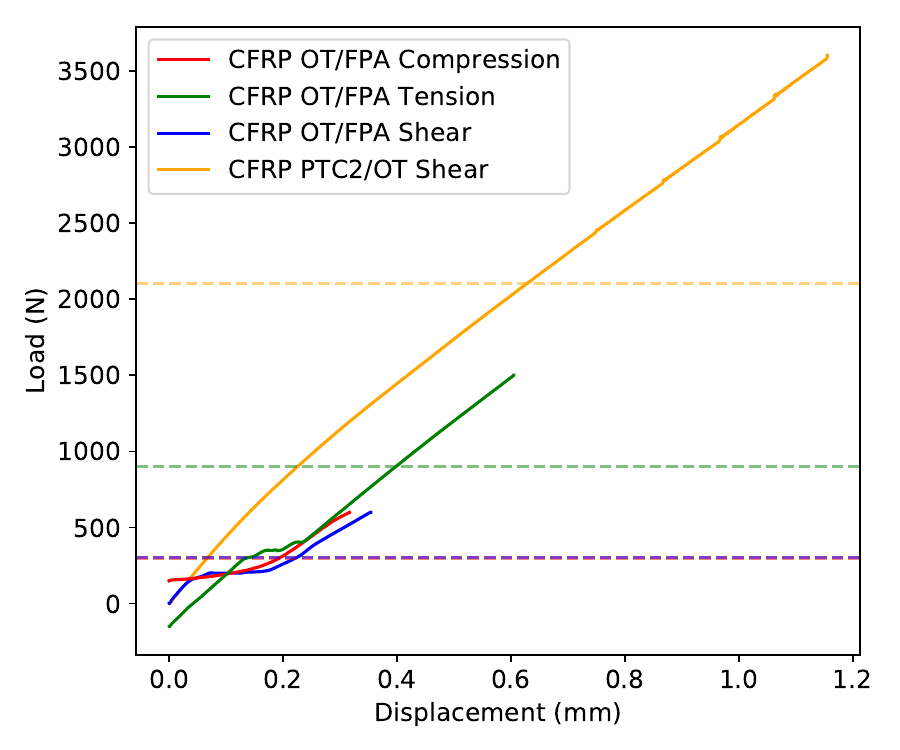}
  \caption{Stress tests of the CFRP PTC2/OT truss in shear and the CFRP OT/FPA truss in all three configurations. After some initial settling, all tests show the trusses remain in the elastic regime as predicted, exceeding the loading requirements set in Table \ref{tab:truss_req} (horizontal lines). The FPA compression and tension tests start from an offset of +150\,N and -150\,N respectively due to the weight of the testing jig. 
  \label{fig:CFtrusstests}}
\end{figure}

We performed non-destructive tests on each CFRP truss. First, the trusses were thermally cycled in liquid nitrogen five times to simulate stresses that occur during cooldowns. Next, we applied loads to the trusses that were safely in the elastic regime of the struts, based on the strut test data, but beyond the minimum requirements specified in Table \ref{tab:strut_req} to ensure the fully assembled trusses were not susceptible to an unanticipated failure mode that would not be found in individual strut tests. This required six tests in total. Four of the six were conducted in an Instron; the CFRP PTC2/OT tension and compression tests were conducted by hanging and stacking weights on the truss, since this truss would not fit in the Instron in those orientations. The tests in the Instron yielded force vs displacement curves, shown in Figure \ref{fig:CFtrusstests}. The CFRP PTC2/OT truss remained in the elastic regime beyond 3.6\;kN in the shear test on the Instron and successfully supported 4.2\;kN in tension and 10.0\;kN in compression. The CFRP OT/FPA truss remained in the elastic regime beyond 1600\;N, 600\;N, and 400\;N in the tension, compression, and shear directions, respectively. The trusses were inspected for damage after each test and with no notable issues, they were approved for installation.

Finally, it is worth noting that there are two additional CFRP trusses that support the detector readout system described in Section \ref{ssec:design-readout}. These trusses support far less mass and are made with the same 3\;mm and 8\;mm OD CFRP struts as the other two trusses, with only small differences in dimension, and so were not separately validated.

\begin{figure*}[ht]
  \centerline{
    \includegraphics[width=\linewidth]
{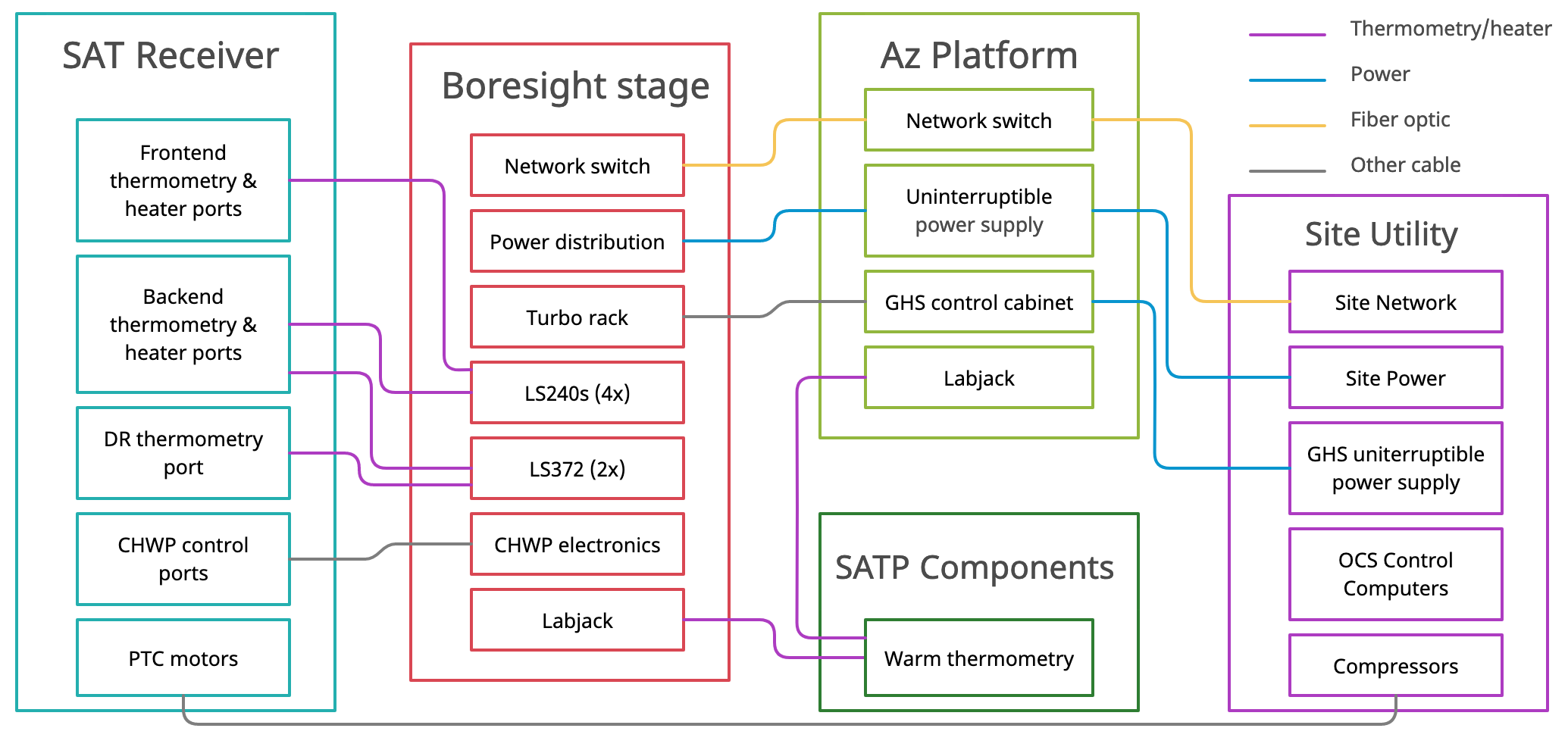}
}
  \caption{Diagram of the SAT housekeeping system and related components. Principal locations of components are identified with different color schemes. Lines denote cable connections between locations. Internal connections, chiefly ethernet and power, are omitted. Also omitted are vendor supplied systems for SATP monitoring such as motors and encoders. More details on the computing structure of the housekeeping system can be found in \cite{Koopman2020}.  \label{fig:HK}}
\end{figure*}

\subsubsection{CFRP truss alignment tests}
\label{sssec:int-cfrpalign}

The CFRP PTC2/OT truss had the same tolerance requirement as the G10 vacuum/PTC1/PTC2 truss and was jigged and glued during manufacture (see Section \ref{sssec:int-g10align}). However, metrology was not performed given the achieved performance of the G10 vacuum/PTC1/PTC2 truss and the relatively loose overall requirements for the OT alignment relative to the receiver shells. 

The CFRP OT/FPA truss sets the positions of the feedhorns relative to the optics. One important feature of the optical design of the SAT is that it provides significant margin in all alignment requirements aside from positioning along the optical axis. The mounting surface for the SAT FPA had to be within $0.20^\circ$ in tip/tilt, 5\;mm in concentricity, and 0.6\;mm along the optical axis to maintain a Strehl ratio degradation of $<1\%$. The truss was attached to an aluminum jig for gluing of the struts to achieve our desired specification. Metrology was performed with a CMM using the SAT-MF1 CFRP OT/FPA truss with the focal plane mounting plate attached and the truss bolted to the OT before the lenses were installed to allow the CMM to access the relevant surfaces. The measurements found the truss to be within tolerance.

\subsection{Magnetic Shielding}
\label{ssec:design-mag}

    The TES detector and readout architecture used in the SATs is susceptible to interference from fluctuating magnetic fields. During operation these can come from the scanning of the telescope through Earth's magnetic field, the drive motors of SATP, and the CHWP bearing and motor. The SAT receivers incorporate multiple levels and types of shielding to maximize attenuation of external fields at the location of the FPA (see Figure \ref{fig:SATX}). Designs were motivated by laboratory measurements of detector and readout sensitivity, simulations, laboratory tests of shielding materials, and previous magnetic shielding designs~\citep{Vavagiakis2020, Huber2021, Connors2022, Ali2020}.
    
    The first shielding layer is composed of Amuneal\footnote{Amuneal Mfg. Corp., Philadelphia, PA 19124} Amumetal 4\;K (A4K) material and comprises a cylinder that attaches to the PTC2 stage and wraps around the OT and FPA volumes with a hole for the optical beam and a lid at the other end to allow FPA access. The second shielding layer is also an A4K cylinder nested inside the first layer and attached to the OT which is thermally coupled to the still stage. A third shield, mounted on the back of the OT, is composed of a cylindrical copper shield around the FPA volume with a 90/10 tin/lead plating on the outer surface to provide a type 2 superconducting shield with a lid at one end of the cylinder to provide access. The lead is included to prevent tin pesting. The impact of penetrations in the A4K magnetic shielding by the 1" diameter copper rods of the heat straps is  minimized with small cylindrical protrusions from each of the magnetic shield lids. More details on the SAT receiver's magnetic shielding can be found in~\cite{Ali2020}. Shielding is also integrated directly into the UFM package, principally in the form of an aluminum shell that composes the outer layer of the UFM package~\citep{Vavagiakis2020, Huber2021}. We report on the achieved shielding factor in Section \ref{ssec:read-mags}.

\begin{figure*}
    \centering
    \resizebox{\linewidth}{!}{
    \includegraphics[trim=200 0 30 0, clip]{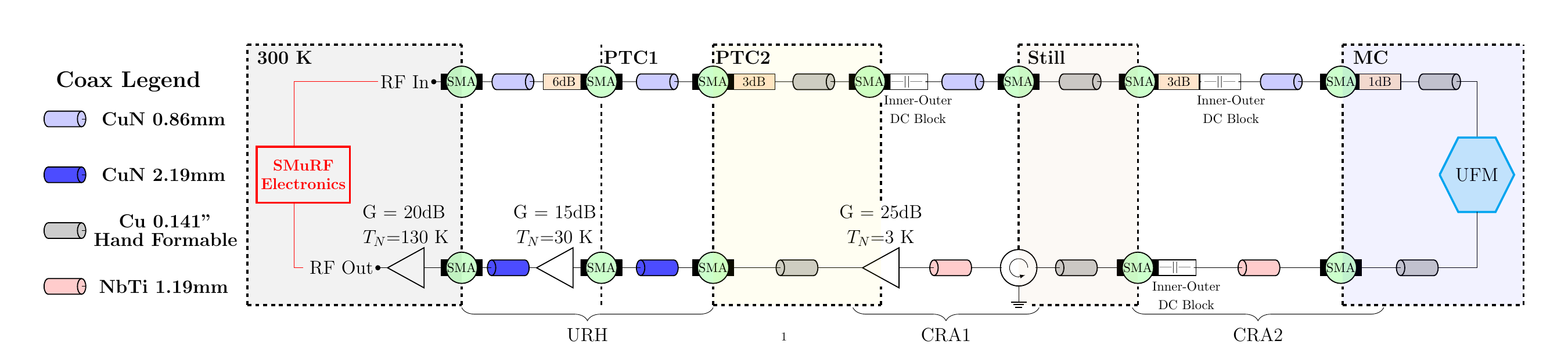}}
    \caption{The RF chain consists of the coaxial cables and components that bring signals from the digital signal processing electronics at room temperature down to the cryogenic multiplexer chips and back up to enable readout of our $\mathcal{O}$(10,000) TES focal plane. The schematic details the components for a single readout chain, with two chains needed for each UFM. The left side of the image is the room temperature SMuRF electronics and on the right side is the connection to the detector arrays at the 100\,mK focal plane. The colored cylinders denote the isothermal and inter-temperature coaxial cables construction material and cross-section with 0.86\,mm CuN (light blue), 2.19\,mm CuN (dark blue), 0.141'' hand formable copper (grey), and 1.19\,mm NbTi (pink). All use solid PTFE dielectrics. The isothermal cables are hand flexed to shape during assembly and are sourced from Mini-Circuits for the SMA to SMA sections and from Centric RF for the 100\;mK SMA to SMP section. The inter-temperature stage coax comes assembled and bent to shape from Coax Co. More details on all of the components in RF chain are given in \cite{Sathyanarayana2020}. Note that in addition to the fixed attenuation, losses in the microwave cabling at temperatures below 4.2K along the “RF In” line reduce the thermal noise within the coax such that it is sub-dominant to other noise terms. }
    \label{fig:SAT_RF_Chain}
\end{figure*}

\subsection{Housekeeping Electronics}
\label{ssec:design-hk}

The SAT receiver requires a suite of diagnostic sensors and heaters to monitor and control the instrument. The most important of these are the cryogenic thermometers, which provide the data necessary for much of the thermal analysis in the following sections. There are sixteen silicon diodes that measure temperatures across the PTC1 and PTC2 stages, and twelve ruthenium oxide (ROX) thermistors that measure temperatures across the still and MC stages. The diodes and still ROXs are read out with four-wire measurements using Lake Shore\footnote{Lake Shore Cryotronics, Westerville, OH 43082} 240 Series Input Modules (LS240s). The five MC ROXs are read out with four-wire measurements using a dedicated Lake Shore Model 372 AC Resistance Bridge (LS372) with associated model 3716 preamp/scanner. In addition, the DR comes equipped with a temperature sensor on each stage, which are read out with a separate LS372 and preamp/scanner.

The ability to apply heat to different temperature stages is important during both testing and normal telescope operation. The DR comes equipped with heaters at the still and MC stages, which are used to set the stage temperatures and to run load curves. A custom heater is located on the focal plane to control and stabilize the temperature of the detectors (see section \ref{sec:int-pid}). These heaters are controlled through the two LS372s. Additionally, several other heaters across all temperature stages were used for in-lab testing to understand the thermal conductivity of the system and were controlled via several programmable PSUs.

\begin{figure}
    \includegraphics[width=1.0\linewidth]{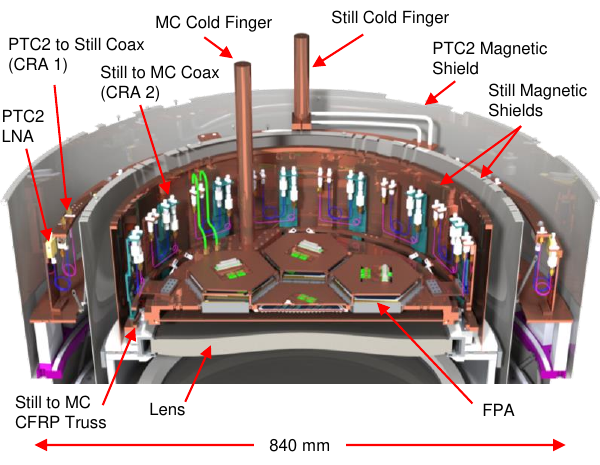}
  \caption{CAD render cross-section view of the mechanical and readout structures from the PTC2 stage to the FPA on the MC stage which includes both cold readout assemblies, CRA1 and CRA2, which provide mounting points for readout components and are the location where the readout cabling bridges the temperature stages. The still to MC CFRP truss is shown in Figure \ref{fig:CFtruss} and an image of the assembly as it appears in the SAT is shown in Figure \ref{fig:backend}. 
  \label{fig:CRA}}
\end{figure}

The thermometry and heater cables enter SAT-MF1 through two Accu-Glass\footnote{Accu-Glass Products Inc., Valencia, CA 91355} 50D2-L100 hermetic feedthroughs. There are four 50 pin twisted pair looms\footnote{Universal Cryogenics, Tucson, AZ 85705} between the vacuum shell and the PTC2 stage designated for: ROX thermometry, diode thermometry, heaters, and a spare. The Spectrum Control Pi filters for the cryostat RF shielding (see Section \ref{ssec:design-mech}) are attached at the hermetic sockets at the output of the cryostat. The ROX thermometry output is connected to a break-out printed circuit board (PCB) which routes the MC ROXs to an LS372 scanner and the still ROXs to an LS240. Manganin cables manufactured by Universal Cryogenics\footnote{Universal Cryogenics Laboratories, Tucson, AZ 85705} with D-sub 50 connectors on one end and MDM-51 connectors on the other route from the vacuum side of the hermetic connector to the PTC2 stage via thermalizing clamps at both the PTC stages. The cables connect to a set of PCBs heatsunk to the PTC2 stage which break-out each MDM-51 connector into 12 4-pin Glenair\footnote{Glenair Inc., Glendale, CA 91201} connectors. Individual sensors and heaters are then routed from the PCB to their installed location on each stage. 

The housekeeping system also monitors and controls a number of other telescope components, including the pulse tube compressors, DR turbo pumps, and gas handling system. The external temperature of the telescope and temperature-sensitive electronics are monitored using approximately twenty thermometers read through a Labjack T7\footnote{LabJack Corporation, Lakewood, CO 80235}. Essential housekeeping systems are powered through an uninterruptible power supply to ensure they can be put into a safe standby mode in the event of power interruption.

The housekeeping system and other devices, such as the CHWP, are managed through the Observatory Control System (OCS), a custom open-source platform designed by SO to manage distributed systems\footnote{GitHub: http://github.com/simonsobs/ocs}. OCS runs across multiple nodes, managing the data acquisition and control for many of the SAT subsystems, as well as  linking them to live-monitoring tools. A detailed overview of OCS and the SAT-MF1 data acquisition network is described in~\cite{Koopman2020}.

The majority of the housekeeping electronics system is located on the boresight stage of the SATP, which holds the SAT receiver. Since the boresight plate is not environmentally sealed, the electronics are protected by a combination of standard and custom NEMA 4X rated enclosures. Power and network connections are provided by electronics at the base of the telescope, which connect to site computing and power outside the ground shields. A block diagram of the housekeeping components is shown in Figure \ref{fig:HK}.

\subsection{Cold Readout Electronics}
\label{ssec:design-readout}

Each SAT has seven UFMs which require a total of 14 input RF chains, 14 output RF chains, and seven cable looms that must traverse five temperature stages with additional mounting points for attenuators, DC blocks, and LNAs. A schematic diagram of the components in the readout chain is shown in Figure \ref{fig:SAT_RF_Chain} with a rendered image of the components surrounding the FPA and a picture of the completed FPA assembly shown in Figure \ref{fig:CRA}. 

The wiring is first routed from the vacuum shell to the PTC2 stage using a ``Universal Readout Harness'' (URH) that is designed to be compatible with all SO receivers. The URH DC connections are brought into the cryostat through welded 50 pin dsub hermetic connectors from CeramTec\footnote{CeramTec, Laurens, SC 29360}. Flexible 36 AWG PhBr cable looms from Tekdata\footnote{Tekdata Interconnections Ltd., Stoke-on-Trent, Staffordshire, ST1 5SQ, UK} carry the TES bias and flux ramp in twisted pairs and amplifier bias signals in twisted triplets through the URH. All looms have Stycast 2850 epoxy backshells and nomex fiber weave. The first loom from the vacuum feedthrough to the PTC1 radiation shield is terminated with a 50 pin dsub connector on the vacuum side and 51 pin MDM connector on the PTC1 end. A PCB is attached to the PTC2 side of the PTC1 radiation shield and breaks out the second stage RF amplifier bias wiring. A second cable runs from this PCB to the PTC2 radiation shield and is terminated with 51 pin MDM connectors on both ends. One set of these looms provides the wiring for the readout of one UFM. The RF coaxial cable construction and RF components mounted within the URH are shown in Figure \ref{fig:SAT_RF_Chain} and described in more detail in~\cite{Sathyanarayana2020}.

The URH can be populated with a variable number of channels depending on the need of the particular receiver. The URH is fully assembled prior to installation using temporary mechanical supports. Once installed, the temporary supports are removed and the weight is transferred onto the SAT shells to minimize thermally conductive paths. Additional details of the URH can be found in~\cite{Moore2022}. 

Two additional assemblies, referred to as the ``Cold Readout Assemblies'' (CRAs), bring the readout wiring from the URH to the focal plane. A schematic representation of the readout chain is shown in Figure \ref{fig:SAT_RF_Chain} and a diagram and photo of the assembly is shown in Figure \ref{fig:CRA}. The CRAs provide transitions between temperature stages, which require semi-rigid low thermal conductivity coax cables integrated into a rigid frame. The CRAs are assembled on the lab bench to allow their installation as two distinct units, one for each temperature transition.  A section of isothermal, hand-formable coax cable is routed from the URH along clamped cable runs and through cutouts in the PTC2 stage magnetic shield to the first cold readout assembly (CRA1). 

CRA1 consists of a copper ring bolted to the PTC2 stage with a CFRP truss offsetting a second copper ring that is coupled to the still stage via a flexible heat strap. Copper brackets are placed around the PTC2 ring which provide mount points for semi-rigid CuNi and NbTi coaxial cables that run between the thermal stages. The brackets also provide a mount point for the first-stage LNAs and a set of co-mounted breakout PCBs. 

Once installed, isothermal coax cables are run from CRA1, through cutouts in the still stage magnetic shielding, to a second readout truss, CRA2. CRA2 also has two copper rings offset by a CFRP truss; one bolted to the still stage of the FPA truss and one connected to the MC stage via flexible heat straps. CRA2 has copper brackets on both rings to mount the CuNi and NbTi coax cables as well as other readout elements. Finally, isothermal coax is routed from the CRA2 MC stage directly to the UFMs on the focal plane.  

Flexible weave cables from Tekdata route the bias lines, flux ramp lines, and amplifier power from the URH to the UFMs with one cable per UFM. The first set of cables goes from the URH to the PTC2 stage of CRA1 where they connect to a PCB that breaks off the amplifier power. The cables are made with 36 AWG copper wire with 12.9 \micron thick polyester enamel insulation twisted in 18 pairs (flux ramp and TES biases) and 4 triplets (amplifier biases) in a Nomex fiber weave and terminate with 51 pin MDM connectors strain relieved with Stycast 2850 epoxy backshells. The second set of cables runs from the PCB to the connection on the back of each UFM with heat sinks at both the still and MC stages. The cables are made with 38 AWG NbTi with 4-6 \micron CuNi cladding and 12.9 \micron Formvar enamel insulation twisted into 18 pairs (flux ramp and TES biases) and terminate with 37 pin MDM connectors with the same weave and strain relief as the copper looms.

The readout architecture of the CRA1 and CRA2 was designed to integrate radially to allow access to the FPA and all readout components by removing the lids of the magnetic shields. Access can be made without disconnecting any electrical connections, greatly simplifying the installation or swapping of UFMs. 

\subsection{Optical Filtering}
\label{ssec:design-opt}

\begin{figure}
  \centerline{
    \includegraphics[width=1.0\linewidth, trim=11 0 0 0,clip]{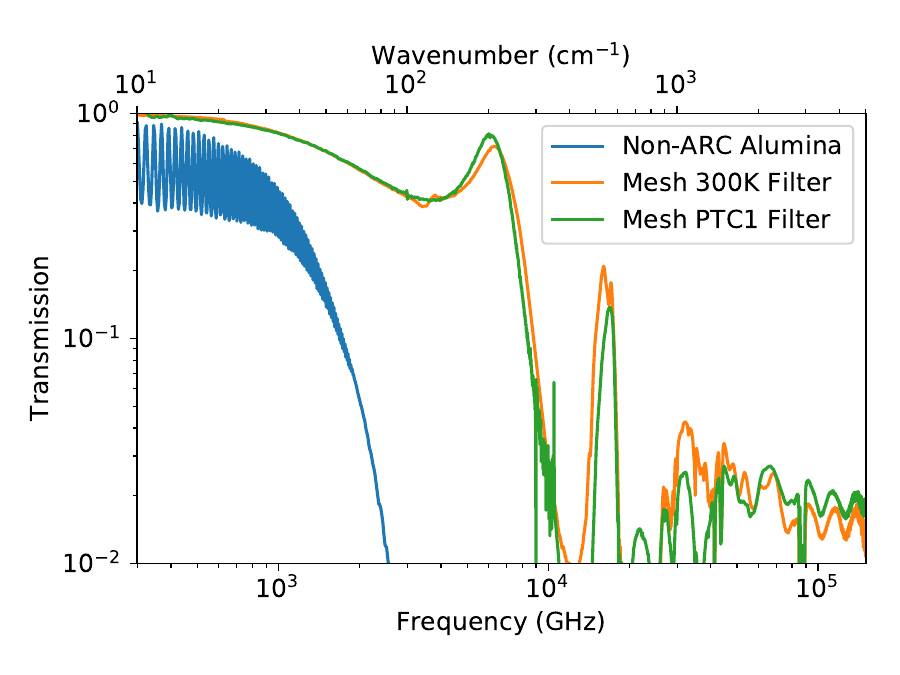}
}
  \caption{Transmission spectra above 10\;cm$^{-1}$ of the IR blocking filters used in the testing phase of SAT-MF1. The two double sided metal mesh filters had 15\,\micron{} grid spacing and 4\,\micron{} substrate thickness. The alumina sample filter has no anti-reflective coating (ARC) but is of a similar type and thickness to the filters used for testing in SAT-MF1. Measurements were made with a Fourier transform spectrometer (FTS) at room temperature at Cardiff. 
  \label{fig:irfilters}} 
\end{figure}

\begin{figure}
  \centerline{
    \includegraphics[width=1.0\linewidth, trim=11 0 0 0,clip]
{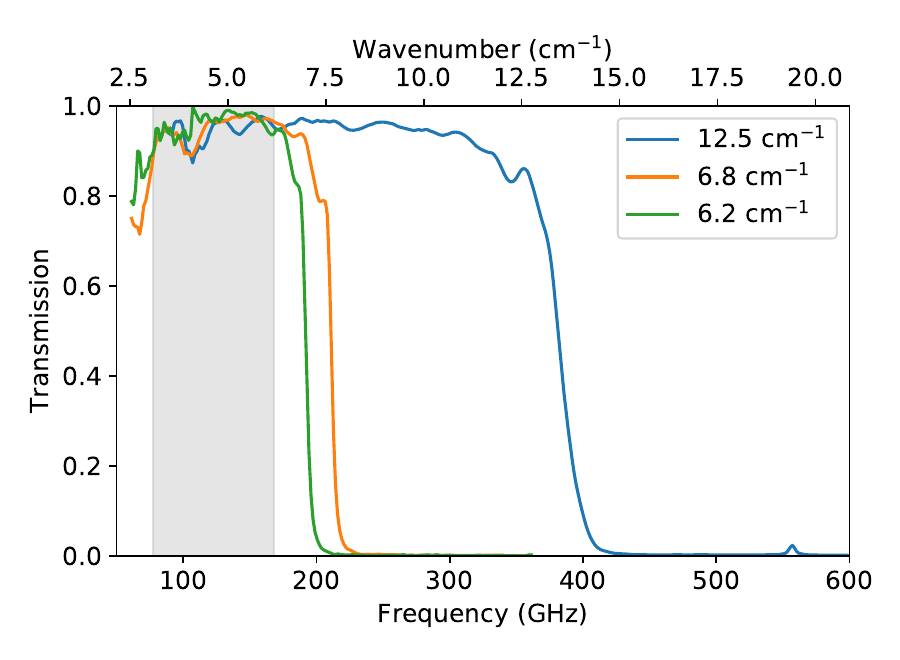}
}
  \caption{Transmission spectra of the three SAT-MF1 LPE filters with the grey box denoting the target bandwidth. Measurements were made with a FTS at room temperature at Cardiff. \label{fig:lpefilters}}
\end{figure}

The filter stack rejects infrared (IR) radiation with minimal in-band attenuation or emission from filtering elements. The wide FOV of the optical design results in a rapidly diverging beam on the sky-side of the still stage aperture stop, motivating a compact thermal filtering chain to minimize the diameter of optical components. Additionally, the CHWP, which is mounted to the PTC1 stage, has a maximum diameter of 50.5\,cm  based on commercially available sapphire plates. As such, the 17\;mm thick, $\approx10$\,kg, spinning CHWP could be no further from the stop than approximately 40\,mm, tightly constraining potential filtering and mounting configurations between the CHWP and the stop. The entire filter chain and CHWP from the aperture stop to window fits in a vertical space of 266\,mm as shown in Figure \ref{fig:SATX}. 

The first element in the chain from the sky-side is a 15\,\micron{} thick polypropylene film with an inner-diameter (ID) of 712\;mm placed in front of the window. The gap between this film and the window is purged with nitrogen and then sealed to prevent dust and snow accumulation directly on the window surface as the SAT is not housed in an enclosure. The protective film can be replaced as needed via an access hatch on the side of the forebaffle base. 

Next, the window is composed of a 10\,mm thick sheet of ultra-high molecular weight polyethylene (UHMWPE) with an expanded teflon anti-reflective coating and an ID of 690\,mm. The window thickness was chosen based on vacuum deflection measurements of 12.7\,mm and 9.5\,mm  thick materials in a dedicated test cryostat with the test thicknesses chosen based on simulations and previous measurements made of fielded UHMWPE windows~\citep{QUIET2013}. The window is designed to deflect no more than 70\,mm under atmospheric pressure which determines the minimum distance to the first filtering element behind the window. A 12\,mm thick window is used for testing in labs at sea level to compensate for the increased pressure. A dedicated testbed was used to measure the window deflection and found the sea level version to deflect $<60$\,mm. The pressure at the telescope site is 0.5\,atm and the 10\,mm thick window was also tested prior to use and found to deflect $<60$\,mm, confirming no risk of interference for testing and operations.

The first two filtering elements used in the testing detailed in the following sections were metal-mesh reflective IR filters~\citep{Ade2006} designed to reject the majority of the $>200$\,W of incident power from the window. The filters had a clear aperture ID of 625\,mm and 524\,mm for the vacuum stage and PTC1 stage filters, respectively. Both filters have a grid spacing of 15\,\micron{} patterned on both sides of the 4\,\micron{} thick polypropylene substrate. The first filter was mounted on the vacuum shell behind the window and the second filter was mounted on the PTC1 stage. A reflective aluminum baffle extends from the vacuum shell to below the rim of the PTC1 IR filter mount to block radiation from the vacuum cavity. 

The volume between the window and the PTC1 filter mount was designed to have sufficient space for a radio-transparent multi-layer insulation (RT-MLI) type filter \citep{Choi2013}. While not used in the testing presented here, a 28 layer RT-MLI filter was ultimately deployed in the field in place of both reflective IR filters as it was found to have equivalent IR performance with a lower instrumental polarization response. Details of its implementation are left to a future publication.

Immediately behind the PTC1 IR reflective filter is the first absorptive alumina filter mounted to the PTC1 stage. A second alumina filter is mounted on the PTC2 stage with the CHWP assembly positioned between the two alumina filters. Alumina was chosen as it has a lower frequency cutoff compared to the metal-mesh filters and has the advantage of being thermally conductive, allowing the filters to absorb radiation and transport the power to the stage shell without the filters themselves heating significantly and re-radiating to lower stages. The filters are 3\;mm thick with a 580\;mm ID and 480\;mm ID for the PTC1 and PTC2 stage filters, respectively. A two-layer anti-reflective coating is applied using a combination of mullite and Duroid to produce a reflectivity of about 2\% at 90/150\;GHz and $<1\%$ at 220/280\;GHz~\citep{Sakaguri2022}. The same anti-reflective coating technology is used on the CHWP. An alternative diced metamaterial anti-reflective alumina surface was developed for SO and implemented on SAT-MF2~\citep{Golec2022}. The testing of SAT-MF1 presented in this paper was performed with 3\;mm thick alumina blanks with no anti-reflective coating applied in order to provide a demonstration of the thermal performance of the receiver, which the alumina anti-reflective coating has a minimal impact on. Details of the IR filter transmission spectra are shown in Figure \ref{fig:irfilters}.

The combination of reflecting and absorbing filters effectively blocks IR power from reaching the 1\;K and 100\;mK stages. Three metal-mesh low pass edge filters (LPEs) are incorporated into the optical stack to remove excess out-of-band radiation. The bandpass of the instrument is formed via the feedhorn coupling and an on-chip filter integrated into the detector architecture. Three LPE filters were used to mitigate resonant `blue-leaks' above each of the filter cutoff frequencies.  The three LPEs in the SATs have 50$\%$ transmission at 12.5\;cm$^{-1}$, 6.8\;cm$^{-1}$, and 6.2\;cm$^{-1}$. The first two LPEs are co-mounted in the OT between the first and second lens with an ID of 508\;mm and are the largest format filters of this type produced. The third LPE is mounted at 100\;mK between the third lens and the FPA with an ID of 427\;mm. Details of the LPE filter transmission spectra are shown in Figure \ref{fig:lpefilters}. 

\subsection{CHWP}
\label{ssec:design-chwp}

The CHWP is part of the optical stack and is located at the PTC1 stage front-end (see Figure \ref{fig:SATX}). A detailed description of the CHWP system can be found in~\cite{Yamada2023} and \cite{Sugiyama2024}. The CHWP sapphire stack is mounted to a magnetic ring, which at cryogenic temperatures is levitated by 53 yttrium barium copper oxide (YBCO) high-temperature superconducting pucks positioned under the ring. Rotation is provided by a combination of small fixed magnets on the rotating component and solenoid motor coils fixed to the PTC1 stage. 

The CHWP system has a number of crucial interfaces that contribute to thermal loading. There are three linear actuated motors that grip the CHWP when the CHWP is mechanically connected to the cryostat, most crucially when the receiver is cooling and warming. The actuators are mounted symmetrically around the CHWP on the vacuum shell and grip the CHWP via three small diameter G10 cylinders, which provide the requisite compression strength while maintaining thermal isolation. The CHWP also has wiring for thermometry, the drive motor solenoids, and the rotation encoder elements. This wiring is routed out of the receiver via two Accu-Glass 15D2-L63 hermetic feedthroughs on the front end of the vacuum shell. The rotation encoder primarily consists of five LEDs and associated photodiodes that provide the spin frequency of the CHWP. The CHWP itself imparts some friction on the system through non-idealities in the magnetic field that induce eddy currents, which can also cause heating. 

\section{Validation and Testing}\label{sec:int}

\begin{figure*}
  \centerline{
    \includegraphics[width=1\linewidth]
{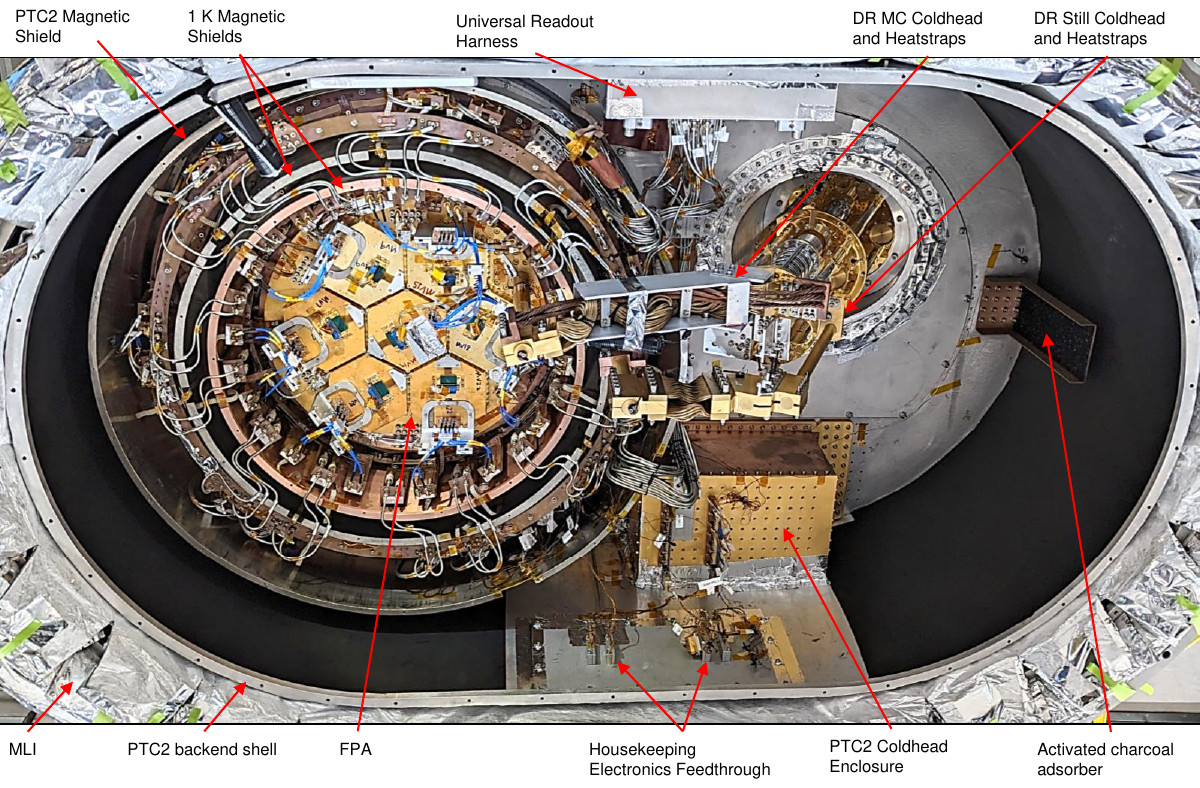}
}
  \caption{Image of the backend access of SAT-MF1 with all lids removed and primary components labeled.  \label{fig:backend}}
\end{figure*}
In this section we describe the validation and testing of the design described in Section \ref{sec:design}. We examined radiative heating, conduction, Joule heating, and the thermal load induced by mechanical vibrations using a series of tests.  The SAT design meets all specifications, with ample margins in terms of cooling capacity.

The SAT-MF1 vacuum and PTC shells were manufactured by Criotec\footnote{Criotec Impianti S.p.A., 10034 Chivasso TO, Italy} and delivered to the University of California San Diego (UCSD) on June 20th, 2019, which started the integration and testing phase of the experiment. SAT-MF1 was delivered with the vacuum, PTC1, and PTC2 shells assembled and the Ruag MLI blankets installed. 

\subsection{Cryogenic validation}\label{ssec:int-cryo}

A series of cooldowns were performed with components incrementally added to allow quantification of the thermal loading contribution from component subsets to assess whether the loading was within acceptable margins. A summary of the expected loading versus the observed loading can be found in Tables \ref{tab:PTCloading-measured} and \ref{tab:DRloading-measured}. The results presented here encompass the assembly of SAT-MF1 over the course of three years up to the installation of a fully populated focal plane with seven UFMs. 

\begin{figure}
  \centerline{
    \includegraphics[width=1.0\linewidth]
{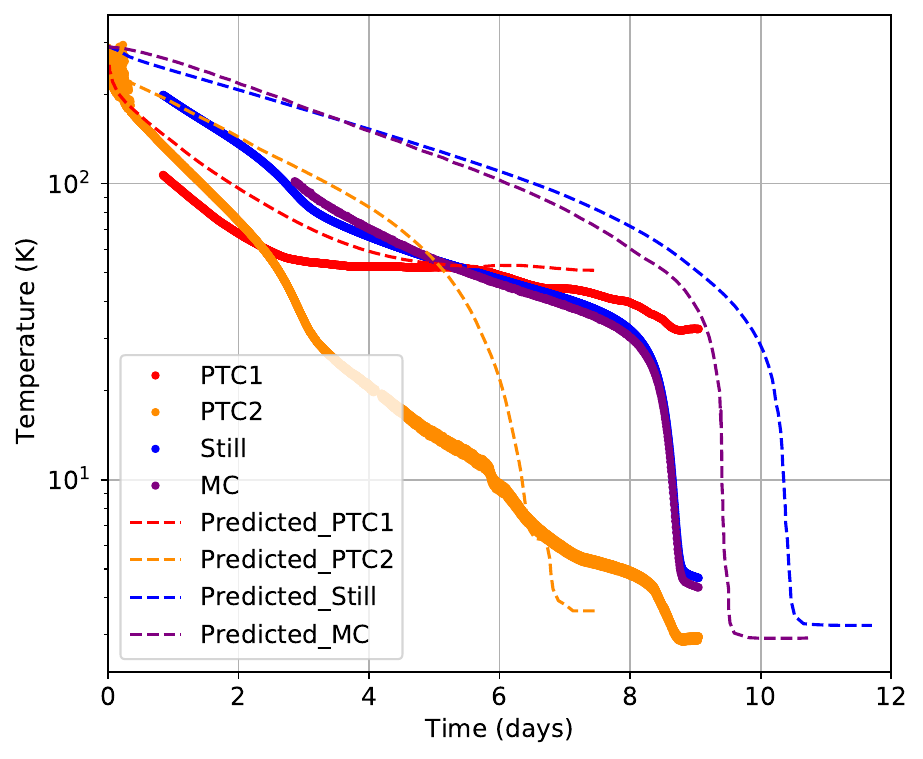}
}
  \caption{Cooldown timeline for the SAT to reach the base temperature of the PTCs. The solid dots are the measured temperature on the cold heads on each of the four stages for a typical cooldown with the correct thermal mass installed. The dashed lines (same color scheme) represent the predictions from \cite{Coppi2018} scaled to the correct mass on the still and MC. 
  It takes about 9 days to get down to the PTC base temperature and 4 more hours to the MC base temperature once the DR circulation was turned on.  
  \label{fig:cooldown}}
\end{figure}

A fully assembled SAT takes approximately nine days to cool from room temperature to below 4\;K on the PTC2 and lower stages, and about 4 hours from turning on the DR circulation to reach base temperature on the MC stage. The cooldown time is 1-2 days faster than predictions scaled from~\cite{Coppi2018} using updated masses of 180\;kg and 30\;kg for the still and MC stage, respectively. More details on the cooldown process are shown in Figure \ref{fig:cooldown}.

\subsubsection{Validation of PTC Stages}\label{sssec:int-ptc} 

The performance of the primary PT420-RM unit was assessed in a dedicated testbed to provide three load curves for use in predicting loading in the receiver. The three data sets are as follows: (1) from before installation into SAT-MF1 at vertical orientation; (2) over two years later after 11 cooldowns of SAT-MF1 in a vertical orientation; (3) during the same test but at $27.5^\circ$ from vertical. Additionally, an extensive set of load curves detailing cooling capacity versus temperature versus PTC orientation relative to gravity was obtained with an identical PT420-RM unit, as detailed in~\cite{Tsan2021}, which allows us to project the two load curve sets of the SAT-MF1 PTC unit obtained in the vertical configuration (0$^\circ$)  to a tilt of 27.5$^\circ$.  The SAT-MF1 testing took place with the cryogenics at this angle, unless otherwise specified. 

\begin{deluxetable}{l | c c }

\tablehead{\colhead{Component} & \colhead{Predicted loading on} & \colhead{Measured loading on}\\[-3mm]
 & \colhead{PTC1(W) / PTC2(W)} & \colhead{PTC1(W) / PTC2(W)}}

\tablecolumns{3}

\tablecaption{PTC1 and PTC2 Stage Thermal Performance\label{tab:PTCloading-measured}}

\startdata
    Base state & 12 / 0.2 & $11 \pm 3$ / $0.10\pm0.08$\\
    URH & 2 / 0.1 & $7\pm3$ / $0.13\pm0.08$\\
    Optical & 5 / 0.1 & $8\pm3$ / $0.17\pm0.08$\\
\enddata
\tablecomments{Predicted numbers include combinations or subsets of values presented in Table \ref{tab:loading}. Base state includes thermal loading from radiative, support, housekeeping cabling, and RF shield sources while the URH loading encompasses all thermal loading associated with the readout cabling.  Measured numbers show loading on the principal PTC, they do not include the additive heat load pulled by the DR PTC. The PTC1 stage excess with the URH is likely due to penetrations in the MLI that are not modeled. The excess was anticipated and fits within design overheads. The possibility of this excess in the PTC1 filter is discussed in Section \ref{ssec:int-filters}. The PTC1 alumina filter performance is inline with expectations and prevents the excess from propagating to lower stages beyond the elevated PTC1 stage temperature. }
\vspace{-0.5cm}
\end{deluxetable}

Two factors impacted our ability to extract the loading imparted on the PTC stages. First, the lab environment was at the local ambient temperature which can fluctuate by over 10\;K between cooldowns, which results in 10\% level drifts in the loading on the PTC1 stage. Second, once the DR is installed, the DR PTC is thermally coupled to the receiver PTC stages while also providing the pre-cooling of the helium for the DR circuit. The loading from the helium circulation is unknown, adding an unquantified and variable loading source to both PTC stages. The values obtained refer to the loading observed on the primary PTC unit. We do not attempt to estimate any loading that is removed from the PTC unit in the DR assembly which is thermally coupled to both PTC stages, though not as tightly as the principal unit. As such, the loading values produced should be interpreted as a lower bound on the total loading. 

By examining the loading using three sets of load curves, multiple loading set points with strategically placed heaters, and COMSOL thermal gradient analysis (see Figure \ref{fig:comsol}), we were able to extract reliable loading estimates that informed whether we were meeting our cooling objectives or that could be used to identify problematic components. A 3\,W uncertainty on PTC1 loading estimates and a 0.08\,W uncertainty on PTC2 loading estimates serve as conservative upper bounds on the values we obtained during our testing. 

We performed a dedicated cooldown to assess the base state loading on the SAT. The following components were installed: the PTC, the DR, the PTC1 and PTC2 shells with penetrations sealed with blank aluminum plates, the MLI blankets including temporary sections to cover the plates, and the housekeeping electronics consisting of six diode thermometers on each stage as well as heaters placed to examine thermal gradients from set loading conditions. The measured total loading is $11 \pm 3$\;W and $0.10 \pm 0.08$\;W on the PTC1 and PTC2 stages, respectively, which are in line with our predictions (see Table \ref{tab:PTCloading-measured}). We also examined thermal gradients by using heaters to apply the expected loading from various components and found the thermal conductivity across the shells to be satisfactory.

\subsubsection{Validation of the DR and DR stages}\label{sssec:int-dr}
\begin{figure}
  \centerline{
    \includegraphics[width=1.0\linewidth]{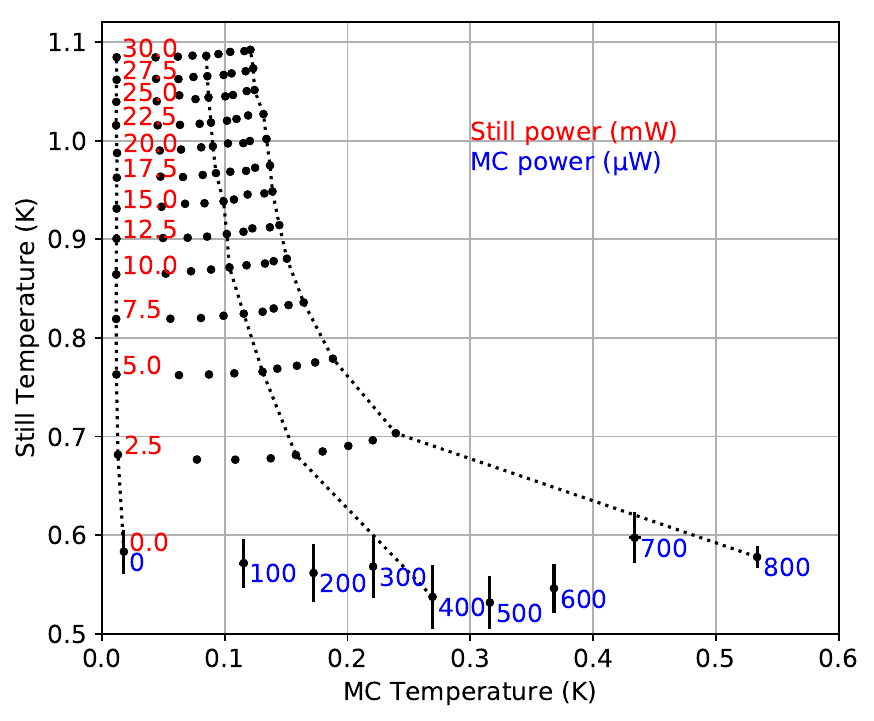}
}
  \caption{DR load curve at $27.5^\circ$ installed but with no components mated to the cold heads. For each still power (red numbers), the lines of dots correspond to temperatures reached at the MC power (blue numbers). The DR PTC2 stage temperature was approximately 3\,K. 
  Power was applied in steps from 0-800\;$\mu$W on the MC stage and 0-30\;mW on the still stage. The same MC set points were used for each still loading row. Error bars are determined by temperature fluctuations over the stable period at each set point (typ. $>10$\,min.) with pronounced variation in points with zero still loading. Dashed lines link three sets of points that share the same MC power, to guide the eye. By performing similar load curves in subsequent cooldowns and examining the shift in temperatures, we are able estimate the additional loading from sub-components on the DR stages. \label{fig:DRLC}}
\end{figure}

The DR was initially cooled in its delivered cryostat using standard piping and hoses to connect components. The DR was charged with 20\,STP liters of $^3$He, which provided a baseline cooling performance of 548\,$\mu$W at 100\,mK with 30\,mW applied to the still stage and with 2.5\,m long condensing and return hoses from the DR and turbo rack to the GHS. 

During operation, the GHS is located on the azimuth platform with the helium condensing and return lines routed through the boresight and elevation cable wraps necessitating 20\;m long hoses from the DR and turbo rack to the GHS (see Figure~\ref{fig:cryoschem}). The DR was tested in isolation after arrival in Chile to assess its performance after 3.5 years of operation and with the 20\,m lines installed. The DR provided 510\;$\mu$W at 100\;mK on the MC stage with 30\;mW applied to the still stage, denoting a 7\% decrease in available cooling capacity which is consistent with earlier tests in the lab with 13\,m long lines. The DR cooling capacity remained well in excess of the 400\;$\mu$W advertised capacity of the system. 

 We performed load curves when the DR was first installed with no additional components attached over a range of loading from 0-30\,mW on the still and 0-800\,$\mu$W on the MC as shown in Figure~\ref{fig:DRLC} with the PTCs and DR tilted 27.5$^\circ$ away from vertical and the PTC2 stage of the DR at approximately 3\,K. The radiative environment was a blackened shell that provided a uniform environment at approximately 4\,K.  These load curve data are used as a baseline to characterize additional loading from new components that were added in subsequent cooldowns (see Table~\ref{tab:DRloading-measured}).

\begin{figure}
  \centerline{       
    \def\big{\includegraphics[width=1.0\linewidth, trim=11 0 0 0,clip]{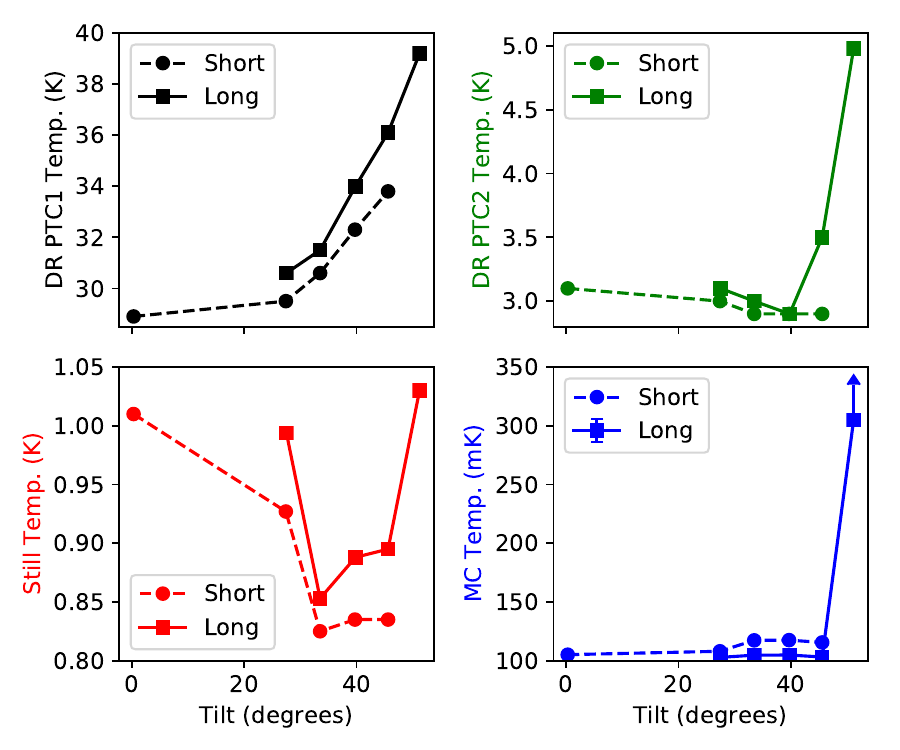}}
    \def\little{\color{black}\fbox{\includegraphics[width=0.24\linewidth]
{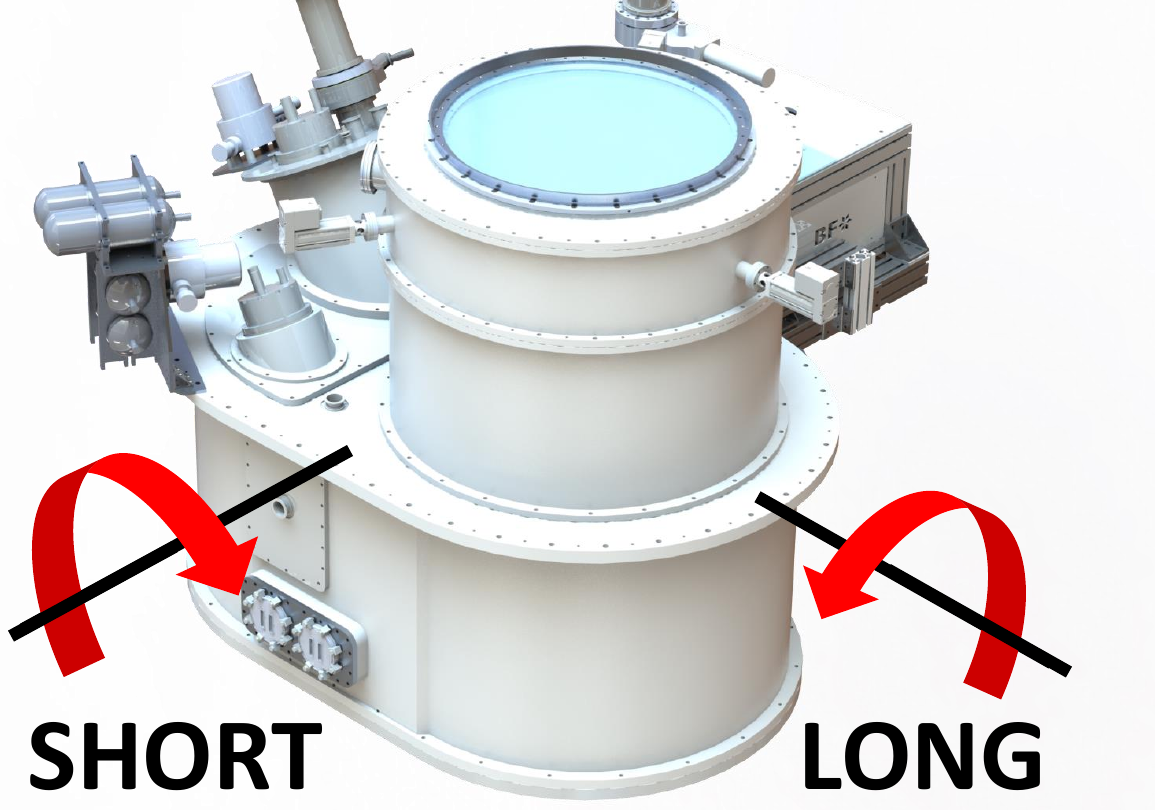}}}
\stackinset{l}{150pt}{b}{35pt}{\little}{\big}
}
  \caption{Temperature response of the PTC integrated into the DR (black and green) and of the DR cold heads (red and blue) over six tilt angles. The tilt angle shown is that of the PTC and DR cold heads relative to vertical, not the SAT optical axis. The Still and MC stages had a constant power applied during testing of 15\;mW and 500\;$\mu$W, respectively. Data from rotation around the short (long) axis of the receiver perpendicular to the optical axis are shown with dashed (solid) lines (see inset). The data were obtained from two separate cooldowns. 
  The MC temperature at 51.2$^\circ$ is a lower bound as the temperature had not stabilized due to the increased temperature on the PTC2 stage effectively exceeding the temperature at which the DR circuit can operate stably. 
  \label{fig:DRTilt}}
\end{figure}

The DR was cooled twice in SAT-MF1 with the receiver mounted in two different orientations, allowing us to test the tilt performance of the unit around two axes (see Figure~\ref{fig:DRTilt}). As a reference point, we observed a cooling capacity of 505\;$\mu$W at 100\;mK on the MC stage using 13\;m long hoses and with the PTCs and DR tilted $27.5^\circ$ away from vertical inside SAT-MF1. At four points at identical rotation angles in the two cooldowns, we see an average change of +0.05\,K on the still stage and -10\,mK on the MC stage. The 27.5$^\circ$ point is the same orientation for both tests suggesting the variation is due to some systematic differences between the cooldowns, such as ambient temperature, rather than due to a dependence of the DR performance on the rotation axis.  

We observed a sharp increase in temperature on all stages when we rotated beyond 45.6$^\circ$ that establishes the operational cryogenic range. The temperature change is due to a decrease in performance of the PTC units resulting in a rise of the DR’s PTC2 stage past the operational range of the DR circuit. The increase in temperature of the PTC is consistent with the cooling capacity falling at large tilt angles as detailed in~\citet{Tsan2021}. We also note that the still stage temperature appears to rise at lower tilt angles and is generally more variable with tilt angle than the MC stage. 

\fboxrule=1pt\relax
\fboxsep=0pt\relax

\begin{deluxetable}{l | c c }

\tablehead{\colhead{Component} & \colhead{Predicted loading on} & \colhead{Measured loading on}\\[-3mm]
 & \colhead{Still(mW) / MC($\mu$W)} & \colhead{Still(mW) / MC($\mu$W)}}

\tablecolumns{3}

\tablecaption{Still and MC Stage Thermal Performance\label{tab:DRloading-measured}}

\startdata
    CFRP/CRA & 0.9 / 19 & $<1.0$ / $28\pm10$\\
    Optical & 4.0 / 7 & $5.0\pm0.4$ / --\\
\enddata
\tablecomments{Predicted numbers include combinations or subsets of values presented in Table \ref{tab:loading}. Loading from all four CFRP support trusses, both CRA1 and CRA2 cabling, and the RF shielding are included in the CFRP/CRA results. The still loading is an upper bound as the loading was below our ability to resolve it. Due to several issues with the DR system later in the testing program, we were unable to extract reliable loading for cooldowns that had the MC stage LPE filter and FPA installed.}
\vspace{-0.5cm}
\end{deluxetable}

\subsection{Filter Loading Estimates}\label{ssec:int-filters}

Thermal loading from the optical chain is one of the primary contributions at all temperature stages and there is a significant degree of uncertainty in the model predictions. Assessing the filter thermal performance was thus a high priority in the testing program. Further discussion of the filter stack and simulations of the loading can be found in Section~\ref{ssec:design-opt}. 

The predicted loading values were based on nominal transmission and absorption spectra combined with material conductivity data (see Figure~\ref{fig:irfilters}). Estimates of the measured loading were generated from load curve data described in the previous sections. We also produced measured loading estimates from changes in the observed thermal gradient between the cold head and filter mounting plates after installing the filters. We used COMSOL simulations, as described in Section~\ref{ssec:design-methods}, to find the loading that reproduces the observed gradient which provides an upper bound on the input power. Measured values were produced using a non-anti-reflective coating 12\.mm thick UHMWPE window with a blank aluminum plate mounted in front of the window at ambient temperature, typically between 290\;K and 300\;K. The blank plate provides a more consistent measurement by decoupling from the room. A 10\,mm thick anti-reflective coated window is used at the Chilean site but it is only rated for 0.5\,atm pressure. A summary of the estimated and measured loading is provided in Tables~\ref{tab:PTCloading-measured} and~\ref{tab:DRloading-measured}.

\subsubsection{PTC1 Stage Filter Loading}
\label{ssec:int-ptc1filt}

The optical loading on the PTC1 stage was found to be $8\pm3$\;W which is higher than the predicted 5\;W. The additional loading is likely due to three primary effects. The first source was determined to come from excess IR transmission through the reflective thermal filters. The loading was due in part to both thermal filters having the same 15\;\micron grid spacing on a 4\;\micron thick substrate, allowing matching resonant features with higher transmission. 

The second possible source of excess loading is tied to the uncertainty of the emissivity of the cavity walls inside the receiver as the surface finish of machined aluminum can vary considerably, and the geometry of the cavity itself can approximate a blackbody. A higher effective emissivity than simulated would reduce the effectiveness of the 300K filter placed behind the window, designed to reflect the relatively high emission from the window itself. We implemented a simple mitigation strategy which consists of a polished sheet metal aluminum baffle that extends from the vacuum IR filter mount to the top surface of the PTC1 MLI surrounding the PTC1 reflective filter. The baffle attempts to lower the effective emissivity of the chamber that radiates power onto the PTC1 filter surface. 

The third potential source of excess loading comes from the MLI seam formed around the filter. In the reference cooldown, the aperture on the PTC1 stage was covered by an aluminum plate with 40 layers of MLI that overlapped with the MLI on the PTC1 stage. Once the filter was installed a seam was created with the MLI taped at the edge of the filter mount leaving an area of exposed aluminum and an effective penetration in the MLI that we did not attempt to simulate for the estimated loading. 

The higher-than-predicted loading is not an issue for the thermal performance given that we designed overhead into the system. Subsequent thermal stages are minimally impacted as the absorbing alumina filter on the PTC1 stage is behind the reflective filter, which conducts away the additional power. The excess does lead to a higher PTC1 filter and cold head temperature which we measured as $58\pm2$\;K and $31\pm1$\;K, respectively. The CHWP yttrium barium copper oxide (YBCO) pucks have a strict operational requirement to be cooler than their 70\;K superconducting transition temperature that we achieve with a significant margin.  Otherwise, the PTC1 stage temperature is only important insofar as it influences the other stages. The results presented here were all achieved with the reflective IR filters. However, we ultimately replaced them with RT-MLI in the field as described in Section \ref{ssec:design-opt}, which achieved similar thermal performance.

\subsubsection{Sub-PTC1 Stage Filter Loading}
\label{ssec:int-otherfilt}

The optical loading on the PTC2 alumina filter was estimated using an identical method to that described for the PTC1 filter. The PTC2 stage is notably simpler with a single 3\;mm thick alumina filter to absorb and conduct away thermal loading from the PTC1 stage and CHWP as well as any IR that is not rejected by skyward filters. The measured loading is $0.17\pm0.08$\;W which is slightly higher than predictions based on fiducial temperatures but in line with expectations given the elevated PTC1 filter temperature described in the previous section.

The optical loading on the still stage is somewhat more complex as the primary optics of the telescope are all connected to this stage: three silicon lenses, two metal mesh filters, the aperture stop, and a series of blackened baffles. We measure $5.0\pm0.4$\;mW of increased loading, consistent with predictions. The thermal optical loading on the MC stage was not possible to extract reliably, as discussed in Section~\ref{sssec:int-dr}, but does not appear to exceed predictions and does not impact our ability to operate the detector arrays.

\subsection{Conductive loading} 

Conductive paths through the housekeeping and readout cabling and the mechanical structure are a significant source of loading. The loading from the G10 vacuum/PTC1/PTC2 truss is encapsulated in the quantities derived in Section~\ref{sssec:int-ptc} (see Table~\ref{tab:PTCloading-measured}) and cannot be separated out from other sources given the receiver architecture. Likewise, the contribution from the housekeeping cabling cannot be separated out and is small with respect to other sources. For example, the estimated loading of the G10 truss is 5.3\;W/0.1\;W versus that of the housekeeping cabling at 0.08\;W/0.002\;W on the PTC1/PTC2 stages. Similarly, the RF shield double-sided aluminized mylar material described in Section~\ref{ssec:design-mech} is integrated into the G10 and CFRP trusses and is included in the predicted and measured loading from the trusses.

A dedicated cooldown was performed to check the loading of the URH as it was the first such device installed in an SO receiver. We measured that the URH contributed additional loads of $7 \pm 3$\;W and $0.13 \pm 0.08$\;W on the PTC1 and PTC2 stages, respectively. The PTC1 loading from the URH is higher than predicted most likely due to the fact that we do not attempt to model the impact from slits in the MLI and reduced overlap in the MLI blankets on that stage.  The PTC2 loading is consistent with predictions, which is in line with the source of excess PTC1 stage loading being radiative as the conductive loading of the cabling material dominates the loading estimate on the PTC2 stage. 

The CFRP PTC2/OT truss and CFRP OT/FPA truss were installed next, along with all the readout wiring going from the PTC2 to the MC stages. We measured the combined loading to be $<1$\,mW and $28\pm10$\,$\mu$W on the still and MC stages, respectively, which is in line with predictions. 

\subsection{Electrical loading}

The PTC1 amplifiers are produced by Arizona State University (ASU) with an estimated operational loading of 60\;mW. We examined the impact of powering three of the amplifiers simultaneously and were not able to measure a significant change in stage temperature, as the loading is subdominant to most other loading sources on the stage. The PTC2 amplifiers are produced by Low Noise Factory\footnote{Low Noise Factory AB, Gothenburg, Sweden} with an estimated operational loading of 4\;mW. We powered three of seven units as a confirmation and measured $5\pm2$\;mW per amplifier when connected to the full readout chain, which is consistent with the expected loading.

The operation of the thermometry and detector arrays presents another loading source. We use four wire measurements for our thermometers using both PhBr and Manganin conductor materials with excitation voltages chosen to ensure any dissipation is well below other loading sources for each stage. The loading from biasing the detectors in transition and operating the flux ramp is estimated to dissipate a total of $<2$\;$\mu$W on the MC stage with all seven UFMs operating simultaneously. The bias cables between the PTC2 and MC stages use NbTi as the conductor material which becomes a superconductor at $\sim9.7$\;K resulting in no Joule heating in the cables during operation. 

\begin{figure}
  \centerline{
    \includegraphics[width=1.0\linewidth]
{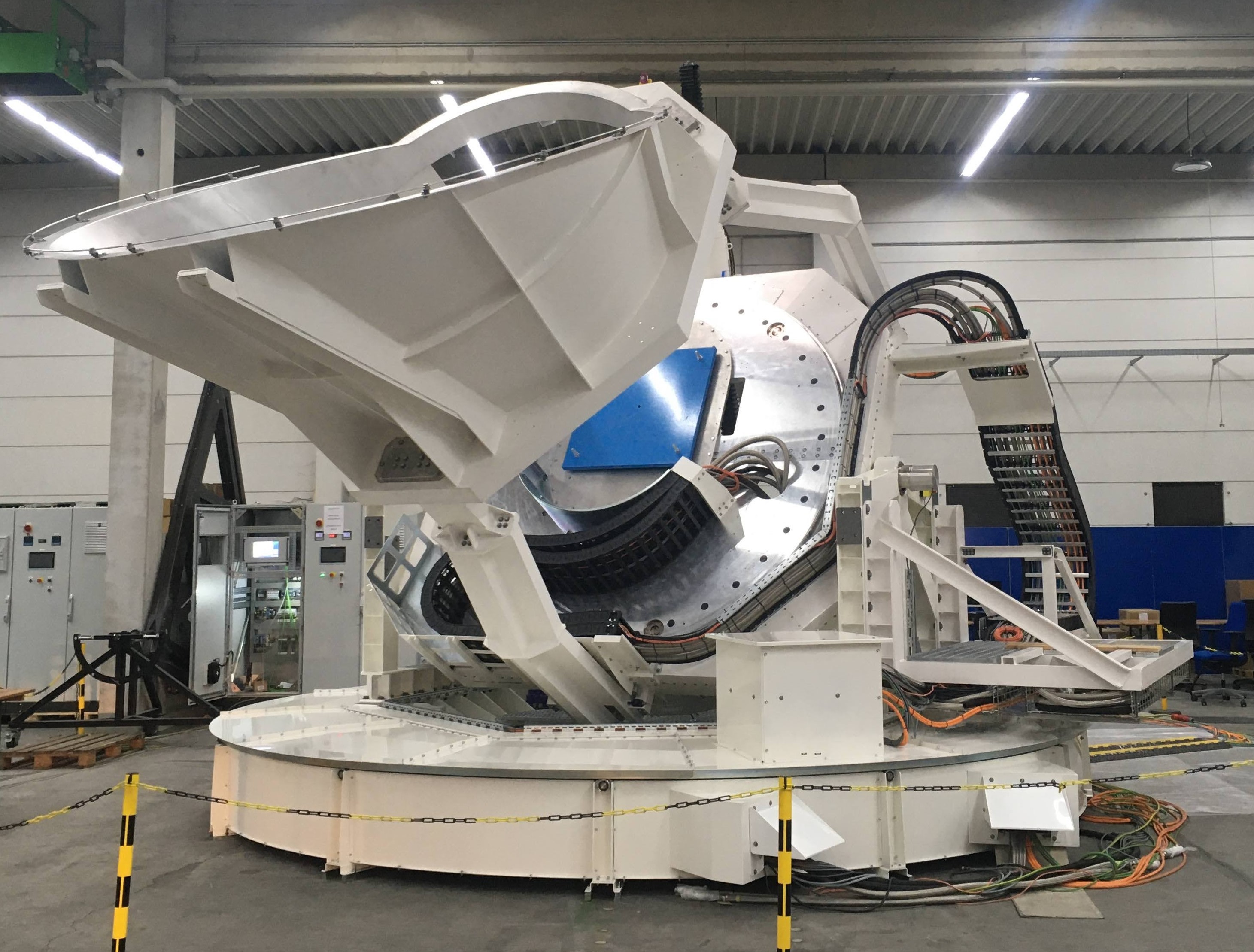}
}
  \caption{The first fully assembled SATP undergoing factory acceptance tests with a steel dummy mass (blue square) installed to simulate the SAT receiver.
  \label{fig:SATP_profile}}
\end{figure}

\begin{figure}
  \centerline{
    \includegraphics[width=1.0\linewidth]
{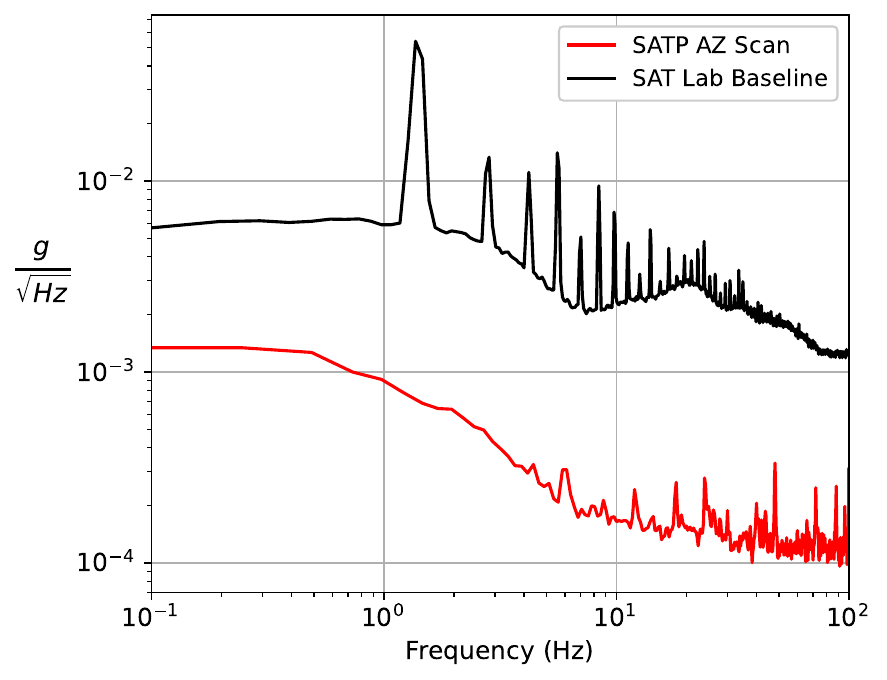}
}
  \caption{
  PSDs of the magnitude of a 3-axis accelerometer attached to the SATP during factory acceptance testing and to the SAT-MF1 mounting flange during a typical cooldown. The SAT Lab Baseline shows the combination of the background vibrational amplitude of the SAT in the lab combined with the 1.4\,Hz vibration associated with the PTCs and its higher frequency harmonics. SATP data were taken during a typical scan in azimuth over 30$^\circ$ at a rate of 1$^\circ$/s with elevation fixed at 50$^\circ$ above horizon and boresight fixed at 0$^\circ$. The SATP PSD was used to investigate potential thermal responses in the SAT. 
  \label{fig:fatpsds}}
\end{figure}

\subsection{Mechanical Resonance Heating}
\label{sec:vibs}

One challenge of operating telescopes at sub-Kelvin temperatures on a moving platform is that system vibrations couple to resonant modes of mechanical elements or wiring, which in turn dissipate the energy as heat. Satellite-based experiments undergo rigorous vibration testing with shake tables, but such testing is typically neither feasible nor necessary for ground based experiments. However, even small amounts of vibrational heating can cause issues at the coldest stage by heating the stage beyond desired ranges or by producing time-dependent heating that can impact detector performance. We have taken a multi-faceted approach to mitigate this risk.

SAT mechanical structures at sub-Kelvin temperatures are designed for vibrational modes to be at a minimum 40\,Hz with a goal to be $>60$\;Hz, where feasible, to push primary modes above the principal low-frequency vibrations of the telescope platform. SolidWorks FEA vibrational mode simulations were used to inform our designs. The principle elements included in the analysis were the CRA1 and CRA2 assemblies, the FPA, and the heat straps for the MC and still stages.

The first validation step was a warm shake table test of the prototype CRA1 assembly and the combined prototype CRA2 and FPA assembly at Quanta Laboratories\footnote{Quanta Laboratories, Santa Clara, CA 95054} to directly measure the vibrational modes at key locations on the structure. The readout components, including wiring and LNAs, were not included in the test to avoid the risk of damage. Mock UFMs made from 3D-printed shells filled with lead weights epoxied in place were used to simulate the mass of the UFMs. However, the plastic material used was likely less rigid than the actual UFMs and differed from the simulations which used solid aluminum bodies of the correct mass as an approximation of the feedhorn blocks which constitute the principle mechanical component of the UFM. The results of the test for the MC stage are summarized in Table~\ref{tab:vibs}. The test found certain elements with resonances close to 40\;Hz that would be straightforward to improve and we ultimately redesigned several aspects of the CRA1 and CRA2 support assemblies to add rigidity on both the still and MC stages. The modifications were incorporated before testing in SAT-MF1. 

The next step in our testing program was to characterize the operational environment of the SATs by measuring the vibration of the platform at Vertex Antennentechnik in Germany. As part of the factory acceptance testing, a model 356B18 triaxial accelerometer\footnote{PCB Piezotronics, 3425 Walden Avenue, Depew, NY 14043} was installed on a SAT receiver mass model that was attached to the SATP during test scans.  Figure \ref{fig:SATP_profile} shows the test setup for the SATP and Figure~\ref{fig:fatpsds} shows the measured SATP vibrational environment.  The vibrational environment was recorded for standard scan patterns and turnarounds and met our highest priority specification for the SATP, summarized as no broadband spectral features with amplitudes $>3\times10^{-4}\;$g$/\sqrt{\text{Hz}}$ at frequencies $>$1.8\;Hz (see Figure~\ref{fig:fatpsds}). Narrow features are potentially a concern but are observed at levels below the SAT-MF1 vibrational background in the lab. 

\begin{deluxetable}{l | c c }

\tablehead{\colhead{Component} & \colhead{Simulated} & \colhead{Measured}\\[-3mm]
 & \colhead{Resonance (Hz)} & \colhead{Resonance (Hz)}}

\tablecolumns{3}

\tablecaption{SHAKE TABLE TEST DATA\label{tab:vibs}}

\startdata
    FPA & 140 & 90\\
    MC Heat strap & 80 & 52\\
    MC CRA2 & 50 & 40\\
\enddata
\tablecomments{Simulations performed in Solidworks finite element analysis (FEA). Measurements are from prototype assemblies that were subsequently modified to increase principal resonances. Several mock parts were used for testing with similar shapes and masses as the actual parts. Higher simulation frequencies are not unusual and are likely caused by several factors including: incorrect loss mechanisms, different materials used for mock parts, and using bonded contacts between components in simulation. }
\vspace{-0.5cm}
\end{deluxetable}

\begin{figure}
    \includegraphics[angle=-90, width=1.0\linewidth]{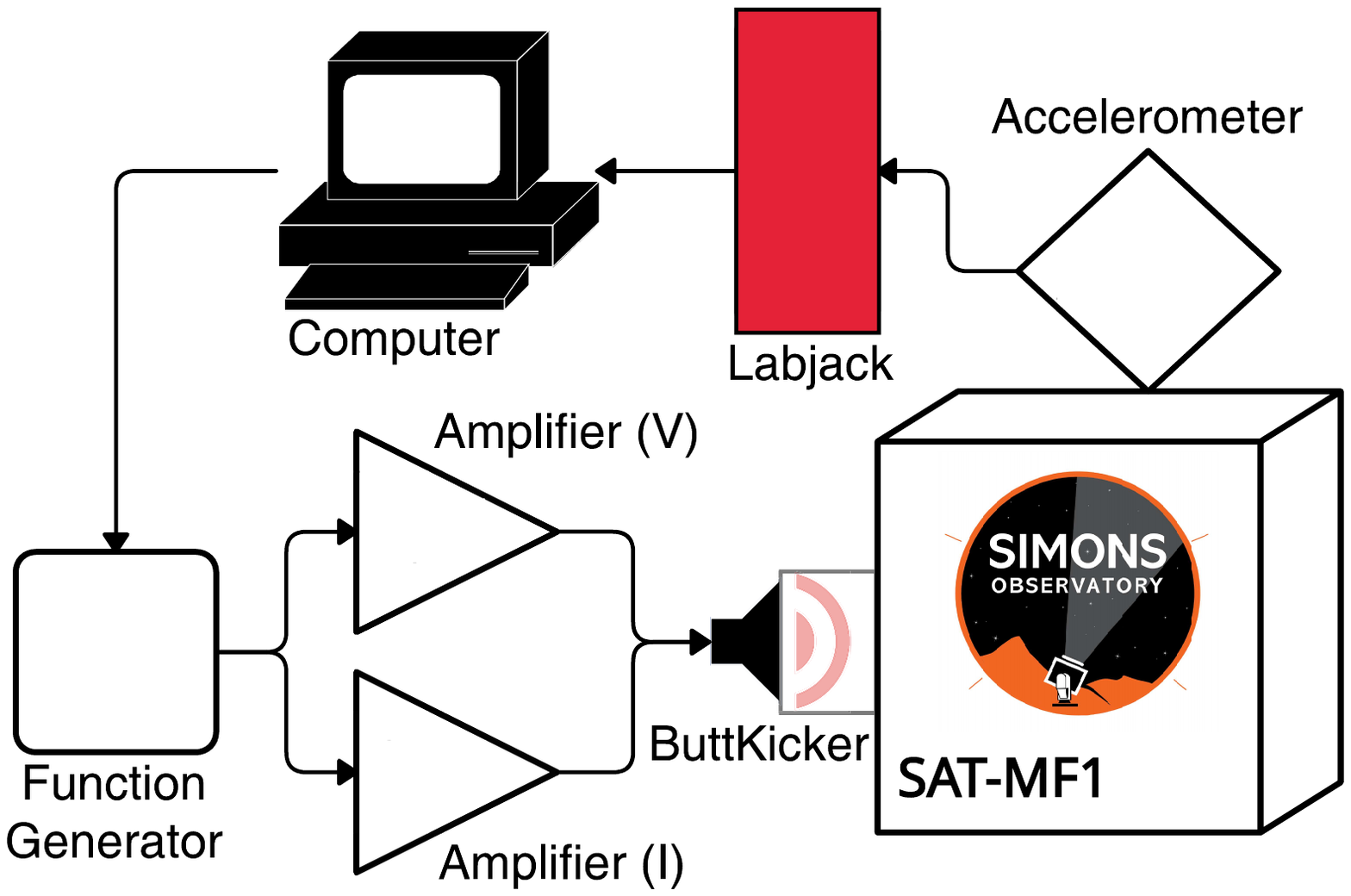}
\caption{Component level diagram for the Buttkicker setup used to conduct vibrational testing on SAT-MF1\label{fig:buttkicker}.}
\end{figure}

 The final component of our testing program involved directly shaking the SAT while at base temperature while monitoring the temperatures at each stage. The results can then be calibrated with the measured SATP vibration environment to examine potential heating scenarios. A vibration test setup was developed for SAT-MF1 based on a similar setup used in the POLARBEAR-2 experiment~\citep{Howe2018}.  
 We bolted a Buttkicker LFE\footnote{The Guitammer Company, Westerville, OH 43086} haptic transducer directly to the bottom flange of SAT-MF1 and drove it with a sine wave from a function generator coupled to two analog amplifiers (see Figure~\ref{fig:buttkicker}). The setup is not as well calibrated as a shake table but it is able to excite vibrational modes at drive frequencies between 10-200\;Hz. The device had to be driven at higher amplitudes with more injected power than is anticipated during normal operations. 
 
 To measure the input vibration amplitude, as well as the background environment, an accelerometer (identical to the one used in the SATP factory acceptance testing) was bolted to the SAT-MF1 window flange with data recorded by a Labjack while the receiver was cold. A frequency sweep from 10-130\;Hz with 1\;Hz steps and was used to identify resonances through heating on the still and MC stages. The step size was chosen based on initial testing that showed expected resonance features had widths of several Hz and due to testing time constraints. It is possible the testing parameters could miss narrow-band resonances. 
 
  A cluster of modes was identified between 38 and 60\;Hz which is associated with the CRA2 MC stage assembly. Multiple resonant peaks are expected given the mounting of the component has several asymmetries which simulations suggest will shift the principal resonant frequency. A peak at 71\;Hz is associated with the heat strap between the DR MC cold head and the FPA based on the measured and simulated resonant frequencies as well as the temperature response of thermometry placed across the MC stage. Several smaller peaks observed around 90\;Hz may be associated with the FPA itself. However, at higher frequency source identification becomes more challenging as the heating can be attributed either to higher order modes of the lower frequencies or primary modes of additional components. 
 
\begin{figure}
  \centerline{
    \includegraphics[width=1.0\linewidth]
{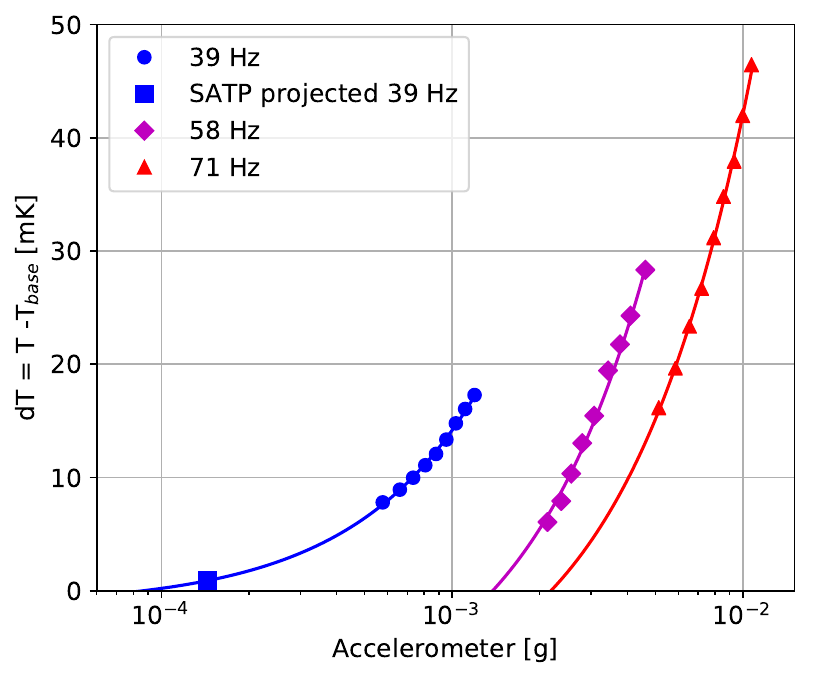}
}
  \caption{The total power of the vibration measured at the accelerometer compared to the change in temperature of a thermometer mounted to the FPA for the three most sensitive vibrational modes associated with the MC stage with linear fits to the data. The projected SATP vibration amplitude is taken from the PSD shown in Figure \ref{fig:fatpsds}. The SATP vibrations at 58 and 71\;Hz are predicted to cause negligible heating. The 39\;Hz mode is predicted to cause less than 1 mK of effective heating during scanning with the time variability a point of future study. The temperature control implemented on the FPA is predicted to be able to compensate for the magnitude and timescale of the effect.}
\label{fig:vibscale}
\end{figure}

The amplitude of the heating response was further probed by varying the input power to the vibration device at three fixed frequencies to understand how the heating scaled (see Figure~\ref{fig:vibscale}). We obtained the magnitude of the spectral line in the accelerometer power spectral density (PSD) [$g^2$/Hz] associated with the vibration drive frequency at each power level for the three resonance modes with the largest temperature response at 39, 58, and 71\;Hz. For each of the axes in the accelerometer, we integrated the peaks over 0.6 Hz frequency range and took the square root to get the power in g. Then we computed the root sum square of all three axes to get the total integrated power [g] and compared ot the rise in temperature of the FPA. A linear fit to each frequency's dT versus the vibrational integrated power was used to extrapolate to the predicted heating at the vibrational power that corresponds to SATP's accelerometer integrated power. We expect the dissipated power to be directly proportional to the injected power down to a floor vibrational amplitude below which the injected power is too low to excite a significant vibrational heating response. 

We established a power for the accelerometer from the SATP vibration PSD by integrating over the same frequency space for each of the target frequencies to identify the expected heating from each mode during the scanning of the telescope. As shown in Figure~\ref{fig:vibscale}, the measured SATP vibration is below the predicted vibrational floor at the 71 and 58 Hz modes, so we do not expect those modes to contribute to heating during operation. The 39\;Hz mode is predicted to excite less than a mK of heating during scanning. An analysis of the scan variability of the SATP vibrational environment and its predicted effect on FPA thermal stability is left for future investigations with the integrated SATP and SAT system. However, we expect the magnitude and timescale of the effect to be such that our active temperature control system described in Section~\ref{sec:int-pid} can readily compensate. 

\subsubsection{Vibrational heating from the CHWP}

During the course of testing, we observed a coupling between the temperature of the FPA and certain CHWP spin frequencies. This was caused by the spinning CHWP exciting vibrational modes in the FPA structure, resulting in up to a 5\;mK increase in the FPA base temperature. The spin frequencies that excited resonances were identified using an accelerometer attached to the SAT receiver, which showed spikes at 96 times the CHWP spin frequency. This multiplicative factor is associated with two symmetry numbers in the CHWP system: the 16 magnetic segments of the rotating CHWP magnetic ring, and the (for thes tests) 48 superconducting YBCO pucks on the static part of the CHWP. The heating is most noticeable when spinning the CHWP from 1 to 2\;Hz, which corresponds to focal plane vibrational resonances at 100 to 200\;Hz. However, since the CHWP can maintain a fixed spin rate to $<0.01$\;Hz accuracy, and the focal plane resonances appear narrow, there are wide regions of spin frequency space to operate in without causing noticeable heating. This allows the CHWP to spin at its desired frequency of $\sim 2$\;Hz without issue. We have also changed the number of superconducting YBCO pucks in the system to 53 such that the number of pucks and number of magnet segments are relative primes of each other to eliminate the observed resonances~\citep{Yamada2023}.

\subsection{FPA Temperature Stability}
\label{sec:int-pid}

Providing a stable thermal environment for the focal plane is important as TES detectors are sensitive to variations in the bath temperature. SAT-MF1 uses a control thermometer combined with a $50\, \Omega$  heater epoxied into a copper mounting block to provide thermal control of the FPA. Superconducting NbTi wires connect the heater to the housekeeping breakout board on the PTC2 stage. We use a Lakeshore 372 temperature bridge and a proportional-integral-derivative (PID) loop implemented in OCS to servo the heater and stabilize the FPA temperature. The temperature measured by the control thermometer is a good approximation for the temperature of the entire FPA, as typical gradients across the FPA are $\leq 1$\;mK. 

\begin{figure}
  \centerline{
    \includegraphics[width=1.0\linewidth]
{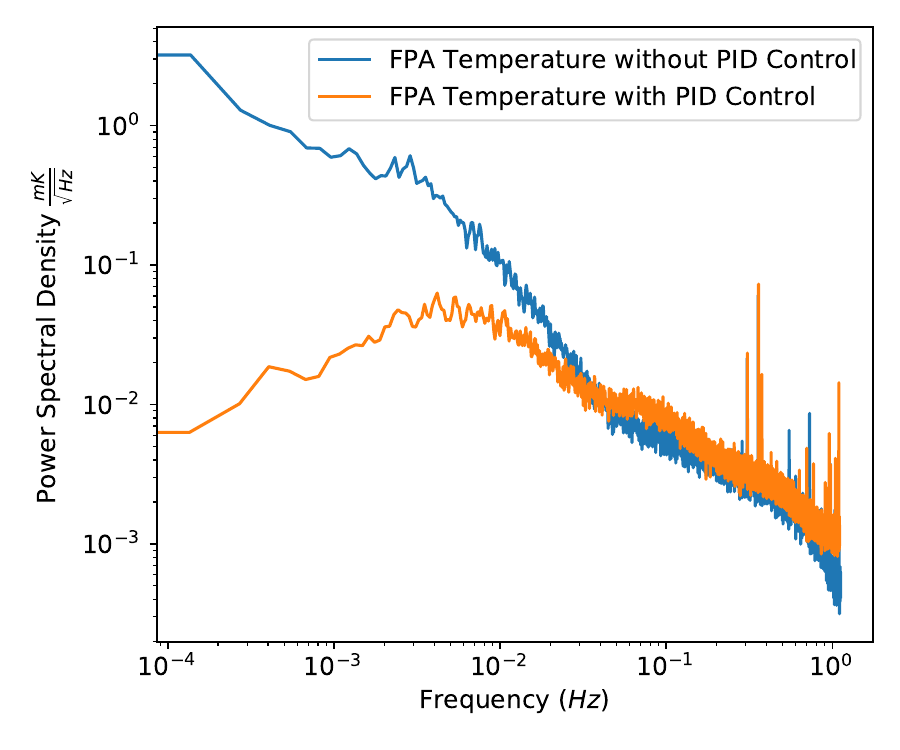}
}
  \caption{Focal plane temperature stability with and without PID control. The `PID control' data were taken at a temperature of 100mK, and the `without PID control' data were taken while constant heating power was applied to the focal plane to produce a similar temperature over approximately 6 hours. A single thermometer sampled at 2Hz was used to record the FPA temperature and provide PID feedback. PID control was optimized for lower frequencies and begins to provide improved stability at periods longer than about 30s. 
  \label{fig:FPA_PID}}
\end{figure}

Through PID control, the SAT is able to control the FPA for an arbitrary length of time with RMS fluctuations of $<7 ~\mu$K . As shown in Fig.~\ref{fig:FPA_PID}, PID control provides improved stability on timescales greater than 30 seconds, eliminating long-timescale effects such as diurnal variations. 

%

\section{Readout and Environmental Testing}
\label{sec:readout}

The performance of certain design elements are best tested by examining the response of the installed detector and readout system. The elements we test include the transmission properties of the RF lines that run through the URH, CRA1, and CRA2, the amplifiers in the system, and the response of the readout and detector system to environmental effects such as magnetic fields. The testing program detailed here covers the implementation of the complete cold readout assembly shown in Figure~\ref{fig:SAT_RF_Chain} for all 14 readout chains in SAT-MF1 as well as assessment of the overall system performance with a detector test unit and a full array of seven UFMs. The detector test unit contains 10 dark detector channels and 50 additional readout channels which include $\mu$MUX resonators coupled to RF-SQUIDs, but without detectors attached. 

\subsection{Readout Transmission Validation}
\label{ssec:read-tranval}
\begin{figure}
  \centerline{
    \includegraphics[width=1.0\linewidth]
{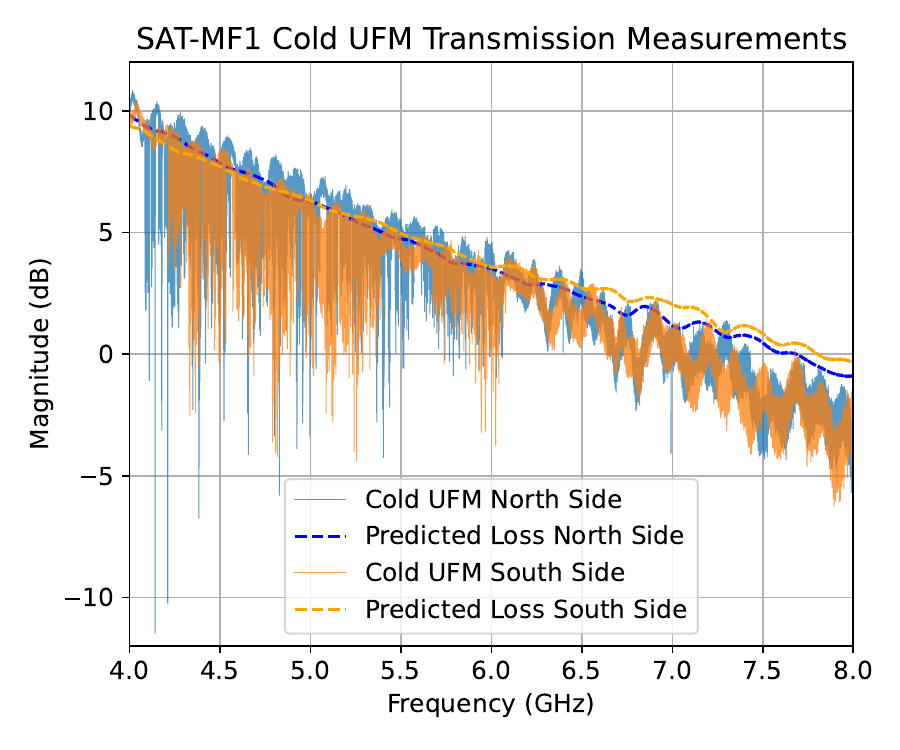}
}
  \caption{Vector network analyzer (VNA) measurements of two readout chains connected to a UFM installed in SAT-MF1. The narrow-band dips correspond to individual detector readout channels. The solid blue and orange lines show the measured data of the two chains, while the dashed lines show the predicted loss from the transmission model which calculates the loss from the RF chain itself. The deviations between the two dashed lines are due to small differences in the components used to assemble each chain (e.g. different coax length, different amplifiers). The measurements are generally consistent with the model over the 4-6\,GHz range with detector tones. Measurements out to 8\,GHz show the extended frequency range capabilities of the readout assembly and the model. The combination of model and measurements have proved useful to check the readout chain performance during testing. Additional attenuation is implemented by the readout electronics system in 0.5\,GHz bins to flatten the response below 6\,GHz (not shown here).} \label{Fig:UFM_VNA}
\end{figure}

The SAT contains seven UFMs with two sets of coax connections on each wafer for a total of 14 RF channels, as well as wiring for TES bias lines, flux ramps, and amplifier power for each channel as described in Section~\ref{ssec:design-readout}. The performance of the coaxial cables and DC wiring for each RF channel was validated at multiple stages of the integration. For this health check, we carried out DC probing and measured the loopback transmission with the VNA of individual chains and compared them to our predicted model. Our model simulates the loss (in dB) of each of the components inside an RF channel and computes the total output loss. The model has been used to quantify the performance of our RF channels and debug problems. Figure~\ref{Fig:UFM_VNA} shows a vector network analyzer (VNA) measurement of a UFM with the complete readout chain compared to the predicted loss. The measurements match well with the predictions over the 4-6\,GHz band that contains the response form the detectors. However, the model does not include effects such as impedance mismatch, leading to some features in the observed signals that are not captured (e.g., standing waves). The performance of the RF channels was tested using a portable Keysight N9916A FieldFox VNA\footnote{Keysight Technologies, Colorado Springs, CO 80907} and compared against predictions at multiple points in the testing program, which confirmed our readout chains performed as expected.

\subsection{Vibration Testing of Detectors and Readout}\label{ssec:read-vib}

Vibrational responses in the readout system and detectors were studied by streaming data with a UFM during the vibrational tests described in Section~\ref{sec:vibs}. During these tests, we were able to see two different types of pickup in the detector timestreams.
In TES-coupled channels, we saw increased 1/f noise where the vibrations caused large changes in bath temperature. In both TES-coupled channels and resonator-only channels, vibrational frequencies close to 11\,Hz also induced a narrow peak in the readout channel PSDs at the driving frequency due to modulation of the RF phase in the warm coax and not anything internal to the SAT. This was addressed by switching to Flexco\footnote{Flexco Microwave Inc., Port Murray, NJ 07865} FC105 coaxial cables in the warm run, which have improved phase stability, and by making sure the cables are well stabilized in the final configuration.

\subsection{Magnetic Field Testing}
\label{ssec:read-mags}

During telescope operation, there are several sources of potential magnetic field pickup. The $\mu$MUX resonators and RF-SQUIDs are sensitive to changes in the local magnetic field, and a magnetic field applied across a TES changes its critical temperature~\citep{Vavagiakis2018}. The two principal anticipated sources of magnetic field pickup are 1) the telescope scanning through the Earth's field and 2) non-uniformities in the permanent magnet on the spinning part (the rotor) of the CHWP. Details of the magnetic shielding in the SAT can be found in Section \ref{ssec:design-mag}.

\subsubsection{Coil Driven Magnetic Fields} \label{Directly injected magnetic fields}

We performed two tests of the magnetic field sensitivity with a field-deployable UFM: once in a test cryostat with no additional magnetic shielding added over the UFM's intrinsic shielding and once in the SAT-MF1 cryostat with all three of the deployment magnetic shields. The magnetic field was generated using a coil of audio cabling wrapped 100 times around a 1\,m $\times$ 1\,m box that was positioned external to the respective cryostat. The field was measured with a Honeywell HMR2300 smart digital magnetometer\footnote{Honeywell Aerospace, Phoenix, AZ 85034} to be uniform to $<10 \%$ within the region spanned by the detectors. A sine wave current with a period of 100\;s was sent through the cabling. This was slow enough to act as an approximately DC field while providing an unambiguous magnetic response in the detectors. The magnetic field at the location of the detector arrays without the cryostat in place was measured using the magnetometer, and a maximum field of $0.8$\;Gauss was applied.

The response of the UFM to the injected magnetic field was determined by measuring the amplitude of the sine-wave in the detectors' time-ordered data, specifically the current through the TES in units of pA. We filter the time-ordered data by: 1) subtracting a linear trend and 2) applying a Butterworth  filter centered on the injected magnetic field modulation frequency. The resulting data product is the response of the UFM to external magnetic fields with units of pA/G of external magnetic field. We observed a clear signal in the test cryostat data. However, no response was seen when the full shielding was present in the SAT-MF1 test so the detector noise level sets an upper bound on the magnetic pickup.

\begin{figure}
    \includegraphics[width=1.0\linewidth]{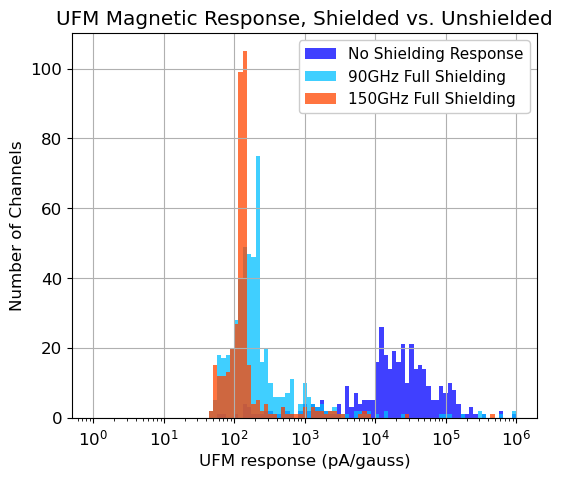}
    \caption{Comparison between the injected magnetic field tests done with a single deployment caliber UFM in a test cryostat with no shielding and in the SAT-MF1 cryostat with full magnetic shielding. The detector population shown was coupled to feedhorns but with a reflective mask at the feedhorn aperture for both tests. The unshielded test (dark blue) exhibited unambiguous responses to the magnetic field in both 90\, and 150\, GHz channels. The shielded test produced no identifiable response above the noise floor of the detector which is used to provide an upper bound on the response level. The noise level varies by frequency band providing a 90\,GHz distribution (light blue) and 150\,GHz distribution (orange). 
    The magnetic field strength is measured by a magnetometer external to the cryostat but at the same distance relative to the solenoid. 
    \label{fig:magnetic_testing}}
\end{figure}

Figure~\ref{fig:magnetic_testing} shows the results of each of the two magnetic field tests. We calculate the lower bound for the total magnetic shielding factor for the entire deployment shielding by dividing the median of the no shielding test response by the median of the detector noise level in the shielded configuration. We measured a median response and standard error of $30,000 \pm 2,000$\,pA/G for the un-shielded test, and an upper bound on the response from the detectors' noise floors in the shielded test of $170 \pm 4$\,pA/G and $130 \pm 3$\,pA/G for the 90\,GHz and 150\,GHz detectors, respectively. We use the 150\,GHz limit to calculate attenuation from the magnetic shielding as it provides the most stringent bound.  We measure a lower bound of total magnetic shielding attenuation of $240 \pm 16$ for a nearly DC field along the optical axis which is the most magnetically permeable axis. 

The maximum variation in the Earth's magnetic field perpendicular to the focal plane during SAT operation can be estimated using our nominal scan strategy, which involves scanning approximately $\ang{45}$ in azimuth at a constant elevation of roughly $\ang{50}$. The amplitude of the Earth magnetic field at the Chilean site that lies in the azimuthal plane is $\approx 0.2$\,G. Thus, we put an upper bound of the injected magnetic field from the Earth to be less than 1\,mG. The upper bound detector signal induced by this injected field (after attenuation due to shielding) is expected to be subdominant to the atmospheric 1/f at the same frequency. At these frequencies ($\sim 10$ mHz), we expect any risidual effect to be minimized by the 8Hz modulation of the polarization signal provided by the CHWP.  

\subsubsection{CHWP-generated magnetic fields}

\begin{figure}
    \includegraphics[width=1.0\linewidth]{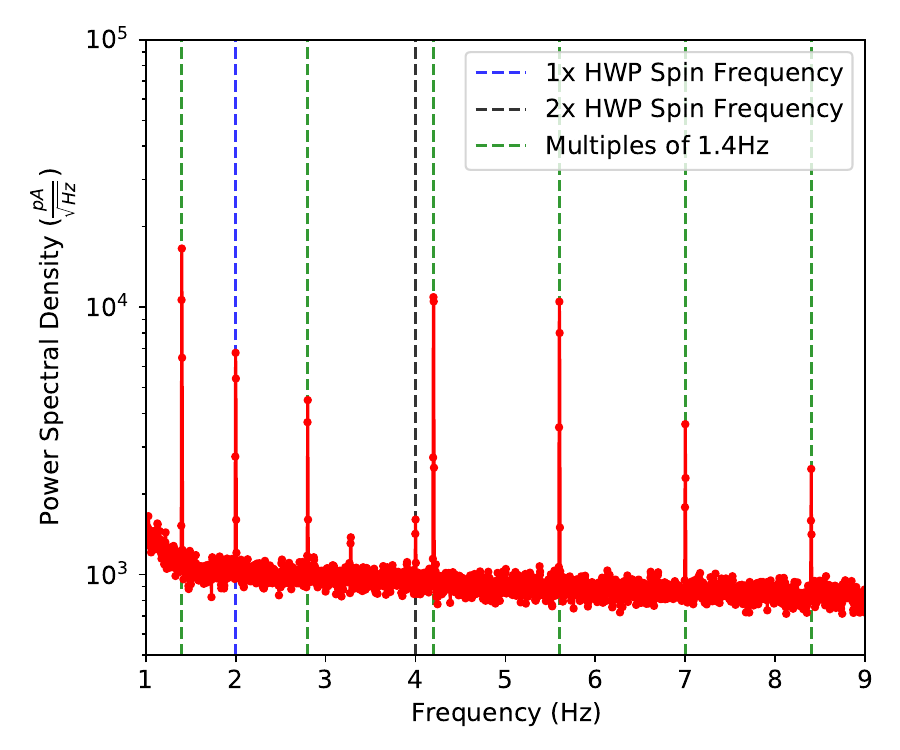}
  \caption{Narrow-bandwidth noise sources seen in a TES-coupled detector from the detector test module from an eight hour integration period in the lab. Testing was performed without the superconducting magnetic shield installed. The pickup at 1x (2\,Hz) and 2x (4\,Hz) the HWP spin frequency is clearly visible. We do not detect pickup at 4x (8\,Hz) the spin frequency which is the modulation frequency of the optical polarized signal being measured. The lines at multiples of 1.4\;Hz are associated with the PTC motor drive frequency and are not seen in resonator-only channels. The unlabelled lines appear in both resonator-coupled and TES-coupled channels at equal strength, and are likely associated with the warm readout system.
  \label{fig:narrow_bandwidth}}
\end{figure}

The CHWP magnetic pickup in dark detector and readout timestreams from the detector test unit is measured by examining the PSDs for peaks at multiples of the CHWP spin frequency. The measurements shown here were done before the still stage superconducting magnetic shield was installed. Clear evidence of HWP-induced pickup at the spin frequency was observed across detectors and resonators at multiple different HWP spin frequencies between $1-2$\,Hz. Pickup at both 1x and 2x the spin frequency was visible when spinning the HWP at $2$\,Hz, see Fig.~\ref{fig:narrow_bandwidth}. The strength of the CHWP pickup across detectors and resonators is consistent with the levels observed in Section~\ref{Directly injected magnetic fields}. This strongly suggests that most of this pickup is due to a variable magnetic field generated by imperfections in the CHWP magnetic bearing. The lines have a narrow bandwidth set by the stability of HWP rotation, which is $\leq$ 10 mHz when PID control is used.
 
We also placed the external magnetometer above SAT-MF1 at an equivalent distance to the CHWP as the FPA and observed a clear magnetic field generated at the spin frequency. These results were observed at three distinct spin frequencies of $1$\,Hz, $1.5$\,Hz, and $2$\,Hz, providing unambiguous evidence that the pickup is associated with the spinning CHWP. The magnetometer recorded magnetic field pickup when the CHWP motor was turned off but the CHWP was still spinning, which indicates that the CHWP drive motor coils and corresponding fixed magnets do not generate a significant magnetic field fluctuation at the distance of the FPA, in line with predictions. 

We expect the primary non-ideality of the rotor magnet to be a dipole, which is confirmed by the fact that the pickup in Fig.~\ref{fig:narrow_bandwidth} is strongest at $1\times$ the spin frequency (2\;Hz), but higher order modes can also be present. Any pickup at higher multiples of the CHWP frequency is more significant for the SAT performance since the incoming polarization signal is modulated at four times the spin frequency. Therefore non-optical signals around this frequency, nominally 8\,Hz, can be transferred to the demodulated data, resulting in potential systematics that would need to be addressed in the analysis pipeline. While the 2x (4\;Hz) line is clearly seen, we are unable to distinguish the 3x or higher order modes from the noise in the eight-hour timestream we used in this analysis. In particular, the 4x HWP signal is not visible above the integrated white noise level.

\section{Summary and Future Directions}
\label{sec:conclusion}

The work presented here focuses on the design of the SO SATs and efforts to bring the SAT-MF1 receiver to a state ready for deployment to Chile, which occurred on July 10, 2023. The final testing program for SAT-MF1 focused on the optical validation of the receiver, the results of which will be presented in a future publication. The optical tests included a measurement of the end-to-end optical efficiency, bandpass, and near-field beam to provide an assessment of the optical performance of SAT-MF1. An additional suite of tests are currently in progress with the SAT fully assembled in Chile as part of the on-site commissioning plan. The testing program and criteria is informed by studies including:~\cite{SOGnF2019, Abitbol2021}. 

The SO SATs utilize a 42\;cm aperture refractive optics system to observe mm-wave radiation with polarization sensitivity. Each SAT will be mounted on its SATP, a three-axis pointing platform, to complete the system. The SAT contains four cryogenic temperature stages with typical temperatures of 45\;K, 4\;K, 1\;K, and 100\;mK, with the final stage housing the TES detector assemblies. The aperture stop, all three silicon lenses, and several filters are positioned on the 1\;K stage to improve the sensitivity of the instrument. Additionally, a continuously rotating half-wave plate is mounted to the 45\;K stage to provide polarization modulation enabling recovery of large angular scale data. A mid-frequency SAT cools $>12,000$ TES detector elements to 100\;mK or below.

SAT-MF1 was assembled and extensively tested in the laboratory prior to its deployment to the Chilean Atacama Desert for observations. We have presented a comprehensive overview of the design, integration, and incremental testing of the first light instrument for SO, SAT-MF1, up to and including a full array of deployable grade UFMs. The testing elements and results presented here include:
\begin{itemize}
    \item Design and testing of the mechanical structure of the receiver including the G10 vacuum/PTC1/PTC2 truss and the two principle CFRP trusses. Our results provided invaluable information on truss design and performance which ultimately yielded components that met our requirements.
    \item Design and testing of the cryogenic system that has successfully cooled one of the largest 1\;K optical assemblies to date in a CMB receiver with over 200\;kg of mass and over 0.26\;m$^3$ in volume.
    \item Characterization of the thermal loading of the receiver and comparisons to loading values derived from a comprehensive set of predictive methods. The results either aligned with expectations or fit within design margins.
    \item Design and validation of both the thermal and RF performance of the end-to-end readout chain assembly for a $\mu$MUX system in a fully assembled CMB camera. 
    \item Assessment of the effectiveness of the magnetic shielding strategy for the receiver and readout system.
    \item A methodology to simulate and assess the vibrational coupling of structures on the still and MC stages of the receiver in the laboratory.
    \item Characterization of the thermal and magnetic interactions of a cryogenic, continuously rotating, half-wave plate system integrated into the receiver with an optical aperture of 480\;mm. 
\end{itemize}

The SAT design demonstrates what can be done with current technology to provide an effective and relatively compact cryogenic system  that can be modified and tailored to other purposes. We are in conversations to develop the SATs for additional applications including for use with kinetic inductance detectors (KIDs) and other devices. 
The development of key technologies and capabilities with the SO SATs will be a valuable reference for future experiments planning to operate with large optical volumes at or below 1\,K, such as CMB-S4~\citep{Abitbol2017, Abazajian2016}.

First light for SO was achieved in late 2023 and on-sky optical tests are being performed prior to beginning routine CMB observations in mid-2024.

\section*{Acknowledgments}
We thank Vertex Antennentechnik GMBH and specifically Christian Bram and Roland Kirchhoff for providing valuable testing and feedback with the SATP. Criotec, Bluefors, and Cryomech have been valuable manufacturing partners that have provided generous feedback and support over the course of the receiver design and testing. We thank the Scripps Institute of Oceanography machine shop for the essential design feedback and production support for many of the custom components that have been machined and integrated into the SATs. Additionally, we thank Joe Saba and Space Dynamics Laboratory for their assistance with detailed FEA of the G10 truss elements. We also thank Aamir Ali for his contributions to the early design and development efforts of the SATs.

This work was supported in part by a grant from
the Simons Foundation (Award \#457687, B.K.). Carlo Baccigalupi acknowledges partial support by the Italian Space Agency LiteBIRD Project (ASI Grants No. 2020-9-HH.0 and 2016-24-H.1-2018), as well as the InDark and LiteBIRD Initiative of the National Institute for Nuclear Phyiscs, and the RadioForegroundsPlus Project HORIZON-CL4-2023-SPACE-01, GA 101135036.  
This work was partially supported by JSPS KAKENHI Grant Numbers, JP23KJ0501, JP23H01202, JP23H00105, JP22H04913, JP20K14483, JP19H00674, and JP17H06134. 
This work was supported in part by the Princeton University Dean for Research Innovation Fund for New Ideas in the Natural Sciences and the Wilkinson Fund. 
This work was supported in part by World Premier International Research Center Initiative (WPI Initiative), MEXT, Japan.  This work was partially supported by JSPS Core-to-Core Program JPJSCCA20200003.  Work at LBNL is supported in part by the U.S. Department of Energy, Office of Science, Office of High Energy Physics, under contract No. DE-AC02-05CH11231. 
Co-funded by the European Union (ERC, POLOCALC, 101096035). Views and opinions expressed are however those of the authors only and do not necessarily reflect those of the European Union or the European Research Council. Neither the European Union nor the granting authority can be held responsible for them. 
This document was prepared by Lauren J. Saunders using the resources of the Fermi National Accelerator Laboratory (Fermilab), a U.S. Department of Energy, Office of Science, Office of High Energy Physics HEP User Facility. Fermilab is managed by Fermi Research Alliance, LLC (FRA), acting under Contract No. DE-AC02-07CH11359. 
Zhilei Xu was supported by the Gordon and Betty Moore Foundation through grant GBMF5215 to the Massachusetts Institute of Technology. 
Jon E. Gudmundsson acknowledges support from the European Union (ERC, CMBeam, 101040169).

\section*{Acronyms and abbreviations}
\noindent 
ARC - anti-reflective coating\\
CFRP - carbon fiber reinforced polymer\\
CHWP - cryogenic half-wave plate\\
CMB - cosmic microwave background\\
CMM - coordinate measuring machine\\
CRA1 - cold readout assembly between the PTC2 and still temperature stages\\
CRA2 - cold readout assembly between the still and MC temperature stages\\
DR - dilution refrigerator\\
FEA - finite element analysis\\
FoS - factor of safety\\
FoV - field of view\\
FPA - focal-plane array\\
FTS - Fourier transform spectrometer\\
GHS - gas handling system\\
ID - inner-diameter\\
IR - infrared\\
LAT - large aperture telescope\\
LNA - low noise amplifier\\
LPE - low pass edge\\
MC - mixing chamber of the dilution refrigerator\\
MLI - multi-layered-insulation\\
OCS - observatory control system\\
OFHC - oxygen-free high conductivity\\
OT - optics tube\\
PCB - printed circuit board\\
PID - proportional-integral-derivative\\
PSD - power spectral density\\
PTC - pulse tube cooler\\
PTC1 - The first temperature stage of the PTC, nominally 45\,K\\
PTC2 - The second temperature stage of the PTC, nominally 4\,K\\
ROX - ruthenium oxide\\
RT-MLI - radio-transparent multi-layer insulation\\
SAT - small aperture telescope\\
SATP - small aperture telescope platform\\
SO - Simons Observatory\\
Still - still chamber of the dilution refrigerator\\
TES - transition edge sensor\\
UFM - universal focal-plane module\\
UHMWPE - ultra-high molecular weight polyethylene\\
$\mu$MUX - microwave multiplexing\\
URH - universal readout harness\\
VNA - vector network analyzer\\
YBCO - yttrium barium copper oxide\\

\typeout{}
\bibliography{SATMF1_InT}
\bibliographystyle{aasjournal}
\end{document}